\documentclass[prx,aps,floatfix,amsmath,superscriptaddress,twocolumn,longbibliography,nofootinbib]{revtex4-2}

\usepackage{amssymb,enumerate}
\usepackage{graphicx}
\usepackage{graphics}
\usepackage{amsmath}
\usepackage{amsthm,bbm}
\usepackage{color}
\usepackage{dsfont}
\usepackage{hyperref}
\usepackage{natbib}
\usepackage{multirow}
\usepackage{natbib}
\usepackage[braket]{qcircuit}

\usepackage{wrapfig, blindtext}
\usepackage[most]{tcolorbox}
\tcbset{colback=yellow!10!white, colframe=red!50!black, 
  highlight math style= {enhanced,
    colframe=red,colback=red!10!white,boxsep=0pt}
}
\usepackage{amsmath}
\usepackage{tikz-cd}
\usepackage{empheq}

\usepackage{graphicx}

\usepackage{anyfontsize}
\usepackage{hyperref}
\usepackage[capitalise]{cleveref}
\usepackage{nameref}
\usepackage{xpatch}
\makeatletter
\xpatchcmd{\@ssect@ltx}{\@xsect}{\protected@edef\@currentlabelname{#8}\@xsect}{}{}
\xpatchcmd{\@sect@ltx}{\@xsect}{\protected@edef\@currentlabelname{#8}\@xsect}{}{}
\makeatother
\usepackage{youngtab}

\usepackage{tikz}
\usetikzlibrary{decorations.pathreplacing}

\usepackage{tikz-cd}

\usepackage{makecell}

\makeatletter 
\renewcommand\onecolumngrid{ 
  \do@columngrid{one}{\@ne}
  \def\set@footnotewidth{\onecolumngrid}
  \def\footnoterule{\kern-6pt\hrule width 1.5in\kern6pt}
}
\makeatother 
\newlength\dlf

\usepackage{amsmath,amssymb}
\usepackage{amssymb,enumerate}
\usepackage{graphicx}
\usepackage{graphics}
\usepackage{amsmath}
\usepackage{amsthm,bbm}
\usepackage{color}
\usepackage{dsfont}

\usepackage{bbm}

\usepackage{lipsum}
\usepackage{lmodern}
\usepackage{tcolorbox}

\usepackage{amsfonts}
\usepackage{graphicx,graphics,epsfig,times,bm,bbm,amssymb,amsmath,amsfonts,mathrsfs}
\usepackage[normalem]{ulem}

\usepackage{subfigure}
\usepackage{dsfont}
\usepackage{braket}
\usepackage{upgreek }
\usepackage{tikz}
\usepackage{natbib}
\usepackage{chngcntr}

\newtheorem{lemma}{Lemma}
\newtheorem*{lemma*}{Lemma}

\theoremstyle{remark}

\newcommand{\bes} {\begin{subequations}}
\newcommand{\ees} {\end{subequations}}
\newcommand{\bea} {\begin{eqnarray}}
\newcommand{\eea} {\end{eqnarray}}
\newcommand{\be} {\begin{equation}}
\newcommand{\ee} {\end{equation}}

\def\>{\rangle}
\def\<{\langle}
\def\Tr{\operatorname{Tr}}
\def\Pr{\textrm{Pr}}

\newcommand{\ignore}[1]{}

\usepackage{amsmath,amssymb}

\usepackage{mathtools}
\usepackage{microtype}

\usepackage{ytableau}
\ytableausetup{smalltableaux}

\crefname{section}{Sec.}{Secs.}
\crefname{claim}{Claim}{Claims}

\begin{document}
\title{ Optimal Qubit Purification and  Unitary Schur Sampling via  
 Random SWAP Tests}

\author{Shrigyan Brahmachari}

\email{shrigyan.brahmachari@duke.edu}
\affiliation{Duke Quantum Center, Duke University}
\affiliation{Department of Electrical and Computer Engineering, Duke University}
\author{Austin Hulse }
\affiliation{Duke Quantum Center, Duke University}
\affiliation{Department of Physics, Duke University}
\author{Henry D. Pfister}
\affiliation{Duke Quantum Center, Duke University}
\affiliation{Department of Electrical and Computer Engineering, Duke University}
\author{Iman Marvian}
\email{iman.marvian@duke.edu}
\affiliation{Duke Quantum Center, Duke University}
\affiliation{Department of Electrical and Computer Engineering, Duke University}
\affiliation{Department of Physics, Duke University}

\begin{abstract}
The goal of qubit purification is to combine multiple noisy copies of an unknown pure quantum state to obtain one or more copies that are closer to the pure state. We show that a simple protocol based solely on random SWAP tests achieves the same fidelity as the Schur transform, which is optimal. This protocol relies only on elementary two-qubit SWAP tests, which project a pair of qubits onto the singlet or triplet subspaces, to identify and isolate singlet pairs, and then proceeds with the remaining qubits. For a system of $n$ qubits, we show that after approximately $T \approx n \ln n$ random SWAP tests, a sharp transition occurs: the probability of detecting any new singlet decreases exponentially with $T$. Similarly, the fidelity of each remaining qubit approaches the optimal value given by the Schur transform, up to an error that is exponentially small in $T$. More broadly, this protocol achieves what is known as weak Schur sampling and unitary Schur sampling with error $\epsilon$, after only $2n \ln(n \epsilon^{-1})$ SWAP tests. That is, it provides a lossless method for extracting any information invariant under permutations of qubits, making it a powerful subroutine for tasks such as quantum state tomography and metrology.

  \end{abstract}

\maketitle

\section{Introduction}
A broad and fundamental class of problems in quantum information science involves extracting and processing information encoded in the quantum state of many qubits. A central obstacle to realizing quantum advantage in these settings is the efficient implementation of the necessary information-extraction operations (e.g., using only a limited set of experimentally accessible, elementary operations).

In many of these applications, the goal is to extract information that does not depend on the order of the given qubits-that is, permutationally invariant (PI) information. This includes, for instance, quantum metrology and state tomography \cite{helstrom1969quantum, holevo2011probabilistic, giovannetti2011advances, giovannetti2006quantum, TomographyHarrow}, where the goal is to use many copies of a quantum state to estimate an unknown parameter or reconstruct a classical description of the state. Another example discussed further below is qubit purification \cite{Cirac}, where many noisy copies of an unknown pure state are combined to obtain one or more copies with higher fidelity. Other important examples of applications include density matrix exponentiation \cite{densitylloyd}, quantum emulation \cite{marvian2016universal}, coherence distillation \cite{marvian2020coherence}, and majority voting \cite{majority}.

 \subsection*{Overview of Results}

\begin{figure*}
\includegraphics[width=10cm]{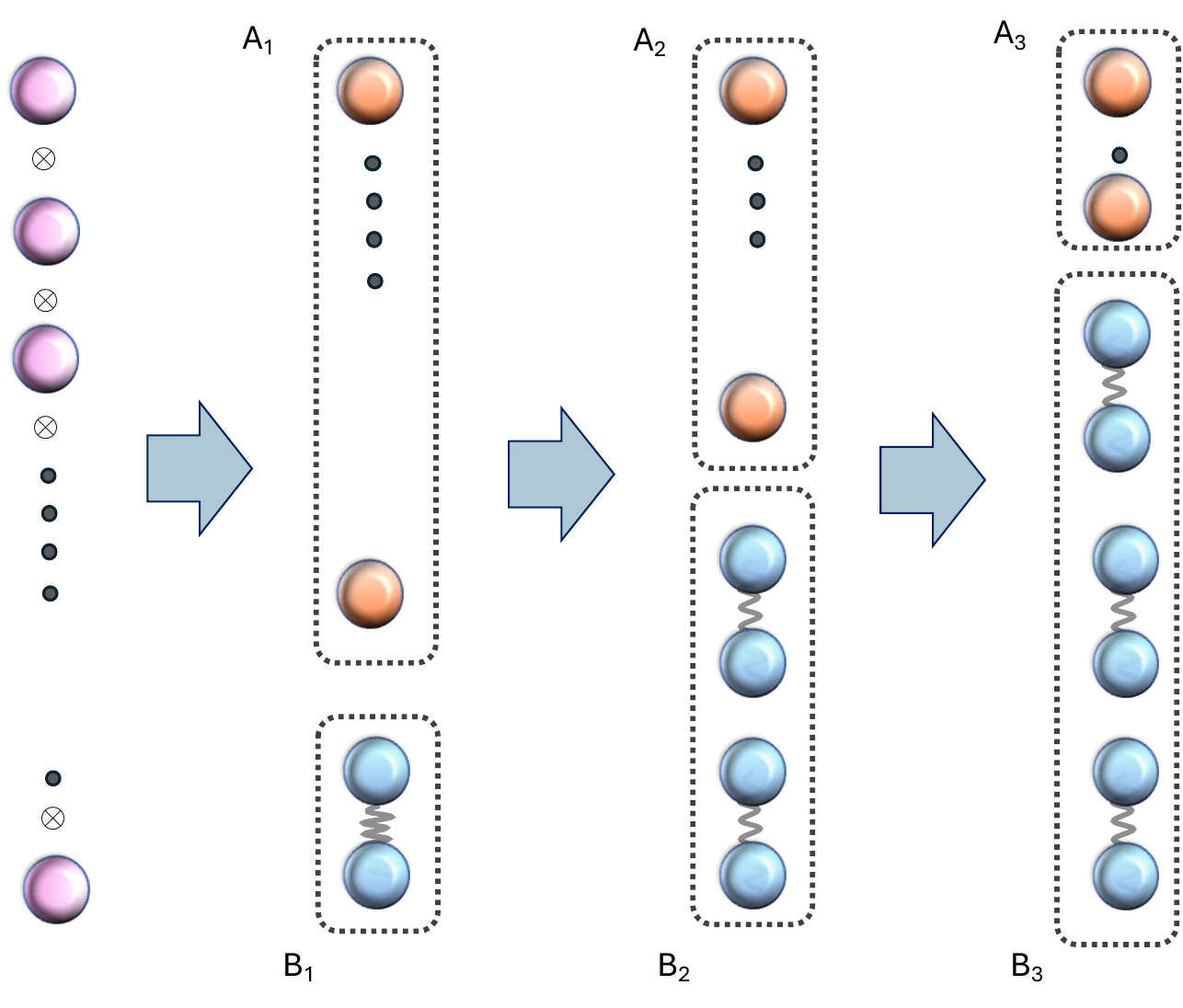}
\caption{\textbf{ Qubit Purification and Unitary Schur Sampling via  
 Random SWAP Tests.} This protocol uses random SWAP tests to detect and isolate pairs of qubits in the singlet state $(|01\rangle - |10\rangle)/\sqrt{2}$. In the figure above, the partitions $B_k$, for $k = 1, \dots$, correspond to the detected and separated singlets. The partitions $A_k$, for $k = 1, \dots$, contain the remaining qubits. At each step $k$, we randomly select a pair of qubits from partition $A_k$, perform a SWAP test on them, and if the outcome projects onto the singlet state, we move that pair to partition $B_k$ and proceed to step $k+1$. Otherwise, if the SWAP test does not detect a singlet, we randomly choose another pair from $A_k$ and repeat the process. We prove that after $ T \ge 2n\ln (n \epsilon^{-1})$ SWAP tests, with probability at least $1-\epsilon$, all singlets in the system have been detected. {Furthermore, for the initial state $\rho^{\otimes n}$ with $\rho = (1-p)|\psi\rangle\langle\psi| + p\mathbb{I}/2$, where $|\psi\rangle$ is unknown, any qubit in $A_k$ will have the highest achievable fidelity given in Eq.~(\ref{opt-Fid}), up to an additional  error of at most $\epsilon$.
}}
\label{Fig}
\end{figure*}

In this work, we show that a simple protocol can efficiently and optimally extract the PI information from systems composed of an arbitrarily large number of qubits.  Remarkably, the protocol requires only the SWAP test, a simple and widely used two-qubit measurement that determines whether a pair of qubits is in the singlet state $(|01\rangle - |10\rangle)/\sqrt{2}$ or in the orthogonal triplet subspace.  
The protocol operates by performing SWAP tests on randomly chosen pairs of qubits. Whenever a pair is found to be in the singlet state, it is discarded, and the protocol proceeds with the remaining qubits (see Fig.~\ref{Fig}). The protocol is \emph{lossless} in the following sense: from its output, one can recover the original input state up to an unknown permutation of the qubits. In this way, the protocol gradually removes entropy from the system while preserving its PI information content.

 We show that for a system of $n$ qubits, after approximately $2 n \ln n$ SWAP tests, a sharp transition occurs: with high probability, no additional singlets are found, and the state of the remaining qubits converges to a fixed state within their totally symmetric subspace. More formally, we compare the output of this protocol with that of the so-called Schur transform~\cite{weakchilds, SchurTransformHarrow, harrowthesis}, a powerful tool in quantum information with broad applications, which, in particular, has been commonly used for extracting PI information. We prove that for PI states, the trace distance between the output of the Schur transform and the protocol based on SWAP tests is bounded by $\epsilon>0$, provided that $T$, the number of SWAP tests, satisfies
 
\be\label{T-bound}
T \ge 2n \ln (n \epsilon^{-1})\ .
\ee

For a system with $n$ qubits, clearly at least $\tfrac{n}{2}$ SWAP tests are needed to interact with all qubits at least once. Therefore, roughly speaking, our result shows that going beyond this minimum by a factor of $\ln n$ is sufficient to implement the Schur transform on PI states, with an error that is exponentially small.

\begin{figure}[!t]
\centering
\makebox[\textwidth][l]{\hspace*{-0.5cm}\includegraphics[width=9.5cm]{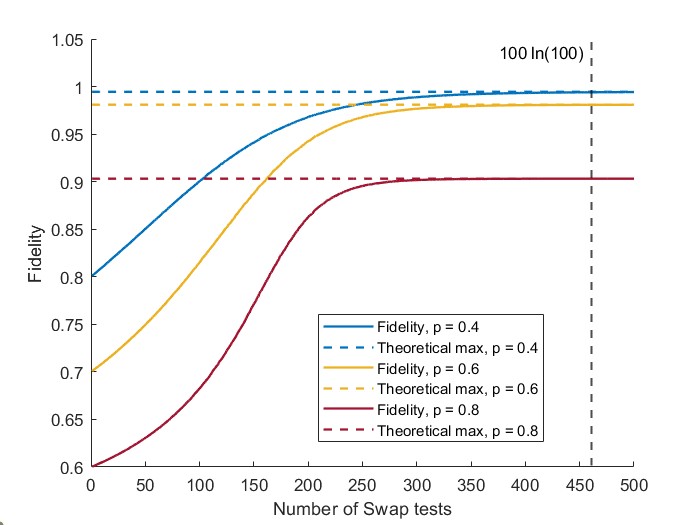}}
    
   \caption{The fidelity of the output qubit of this purification protocol, with the desired ideal pure state, after $T$ SWAP tests. Initially the system contains $n=100$ copies of $\rho=(1-p)|\psi\rangle\langle\psi| + p\mathbb{I}/2$. As the number of SWAP tests grows, the fidelity of the output state converges to the optimal fidelity given by Eq.(\ref{opt-Fid}). The vertical dotted line corresponds to $n \ln n$, which roughly indicates the order of the number of required SWAP tests. See Sec.\ref{Sec:App:Pur} for further details.}
  \label{fig:fidconvergence}
\end{figure}

As an example of applications, we demonstrate that this protocol for state purification \cite{Cirac}. Namely, we apply it to  $\rho^{\otimes n}$, i.e.,  $n$ copies of state $\rho = (1-p)|\psi\rangle\langle\psi| + p\mathbb{I}/2$, where $|\psi\rangle$ is an unknown pure state and $0 <p < 1$, in order to obtain a single qubit with higher fidelity with respect to  $|\psi\rangle$. Cirac et al. \cite{Cirac} have shown how the optimal fidelity can be achieved using the Schur transform, and the fidelity of the output $\rho_{\text{opt}}(n)$ with the desired output state $|\psi\rangle$ is given by 
\be\label{opt-Fid}
\langle\psi| \rho_{\text{opt}}(n)|\psi\rangle =1 - \frac{1}{2n} \, \frac{p}{(1 - p)^2} + \mathcal{O}\left(\frac{1}{n^2}\right)\ .
\ee
Our result then reveals that, up to an arbitrarily small additional error $\epsilon$, the same fidelity can be achieved using $2n\ln (n \epsilon^{-1})$ random SWAP tests, by simply returning one of the qubits that has not been discarded by the SWAP test. See  Fig.~\ref{fig:fidconvergence} for an example, and Sec.~\ref{Sec:App:Pur} for further details.  

As we further explain below, the working principle behind this purification protocol is simple: for PI states, the only relevant information is encoded in the reduced state of the $\mathrm{SU}(2)$ irreducible representations (irreps). Since the SWAP test on each pair of qubits respects this symmetry, the initial state can always be recovered after any number of SWAP tests, regardless of the specific choices of measured qubits and the outcomes. By discarding the detected singlets and continuing the SWAP tests on the remaining qubits, we gradually detect all the singlets in the system. This ultimately yields  a state restricted to the totally symmetric subspace of the remaining qubits, where the SWAP test no longer detects any singlets. In this case, by discarding all the remaining qubits except one, we achieve the optimal purification limit in Eq.(\ref{opt-Fid}).

We note that, prior to this work, Childs  \textit{et al.}~\cite{Childs2025streamingquantum} used sequential SWAP tests to implement qubit purification. However, in their approach, when a SWAP test identifies a pair of qubits in the triplet subspace, one qubit is retained and forwarded to the next level, and the other is discarded. 
  This simplifies the analysis, as the remaining qubits are fully uncorrelated, but it leads to wasted resources and thus suboptimal fidelity as well as suboptimal gate complexity (see Sec.\ref{Sec:App:Pur}).

 We also note that in the context of symmetry-protected quantum computation, Freedman  \textit{et al.}~\cite{freedman2021symmetry} showed in 2021 that the outcomes of random SWAP tests, referred to as triplet/singlet measurements in their work, determine the total angular momentum of a system of qubits.  Building on this, Rudolph and Virmani~\cite{rudolph} showed in 2023 that, starting from qubits in the maximally mixed state, one can, by repeatedly measuring angular momentum in this way, gradually grow a subset of qubits whose joint state is restricted to their totally symmetric subspace, namely, in the maximally mixed state on this subspace. (At each step, one performs random SWAP tests between the previously obtained subset and a new qubit in the maximally mixed state).

 Our results reveal broader applications of random SWAP tests, namely for performing the  Schur transform on PI states, which is useful, for example, in the context of state purification and tomography, and they provide more precise and tighter error analysis, uncovering a 
sharp threshold phenomenon around $T_\ast \approx 2 n \ln n$.  \footnote{Rudolph and Virmani \cite{rudolph} show that the convergence can be achieved to the maximally mixed state on the totally symmetric subspace in $\approx n^2$ steps.  Freedman et al.~\cite{freedman2021symmetry} briefly discuss using random SWAP tests for measuring angular momentum $j$, as a useful subroutine in the so-called symmetry-protected quantum computation. They argue that ``since spin is quantized, the convergence is actually faster than polynomial. Once the number of measurements is polynomially large compared to $n^2$, the convergence in accuracy becomes exponential.''
 }  We also note that, in the special case of $n=3$ qudits, it was previously observed in \cite{li2024optimal} that unitary Schur sampling can be achieved via sequential SWAP tests.

\section{Setup}
Consider a system of $n$ qubits with the total Hilbert space $(\mathbb{C}^2)^{\otimes n}$. We are interested in the properties of quantum states that are independent of the order of the qubits. More formally, for any state $\sigma$, we are interested in the information that can be obtained from the PI state $\mathcal{P}(\sigma)$, where $\mathcal{P}$ is the quantum channel defined by applying a permutation on $n$ qubits, chosen uniformly at random from the group of all such permutations $\mathbb{S}_n$  (See Appendix \ref{App:SU(2)} for further details).  By applying such random permutations, any state $\sigma$ is projected to its symmetrized version $\mathcal{P}(\sigma)$. Hence, in the following, we often restrict our attention to the case of PI states.

As we discuss further below, thanks to \emph{Schur-Weyl duality}, such states can be conveniently characterized using the representation theory of the group  $\mathrm{SU}(2)$,  i.e., single-qubit unitaries with determinant 1, which is widely used in physics (e.g., to describe rotations).  Recall that under the action of $\mathrm{SU}(2)$ symmetry, $U \mapsto U^{\otimes n}$ with $U \in \mathrm{SU}(2)$, the Hilbert space decomposes into subspaces corresponding to inequivalent irreps of $\mathrm{SU}(2)$. These are commonly labeled by the eigenvalues $j(j+1)$ of the squared angular momentum operator (also known as the \emph{Casimir operator})  $J^2 = J_x^2 + J_y^2 + J_z^2$, 
where  $
J_w = \sum_{i=1}^n \sigma_w^{(i)}/2 \quad \text{for } w \in \{x, y, z\}.$ 
The possible values of $j$ are integers $j = 0, \ldots, n/2$ when $n$ is even, and half-integers $j = 1/2, \ldots, n/2$ when $n$ is odd.

Then, under the action of $\mathrm{SU}(2)$ transformations, the total Hilbert space of qubits decomposes as 
\be\label{dec}
(\mathbb{C}^2)^{\otimes n} \cong \bigoplus_{j=j_\text{min}}^{n/2} \big(\mathbb{C}^{2j+1}\otimes \mathbb{C}^{m(n,j)}\big)\ ,
\ee
where  $\mathrm{SU}(2)$ acts irreducibly on $\mathbb{C}^{2j+1}$ with multiplicity 
$m(n,j)={\tbinom{n}{\frac{n}{2}-j}}\times \frac{2j+1}{\frac{n}{2}+j+1}$.  According to Schur-Weyl duality, the group of unitaries generated by pairwise SWAPs of qubits, which is a representation of $\mathbb{S}_n$, is block-diagonal with respect to this decomposition, acts trivially on the subsystem $\mathbb{C}^{2j+1}$ and acts irreducibly on $\mathbb{C}^{m(n,j)}$. Indeed, each $\mathbb{C}^{m(n,j)}$ with a different $j$ corresponds to an inequivalent irrep of this group. In particular, the $(n+1)$-dimensional subspace corresponding to the maximum angular momentum $j = n/2$, also known as the totally symmetric subspace, carries the trivial representation of the permutation group $\mathbb{S}_n$. This subspace includes all pure states that are invariant under permutations, such as GHZ and Dicke states.

It follows that any PI $n$-qubit state $\sigma$, such as  $\sigma=\rho^{\otimes n}$, is block-diagonal with respect to this decomposition and  takes the form
\be\label{dec1}
\sigma=\bigoplus_{j=j_\text{min}}^{n/2} p_j \left(\sigma_j\otimes \frac{\mathbb{I}_{m(n,j)}}{m(n,j)}\right)\ ,
\ee
where $p_j = \Tr(\sigma \Pi_j)$ is the probability that the state $\sigma$ is found in the subspace with total angular momentum $j$, i.e., the subspace $\mathbb{C}^{2j+1} \otimes \mathbb{C}^{m(n,j)}$, $\Pi_j$ denotes the projector onto this subspace, and $\sigma_j$ is a density operator on $\mathbb{C}^{2j+1}$, the irrep of $\mathrm{SU}(2)$ with angular momentum $j$.

\section{Unitary Schur Sampling}

For many applications in quantum information and computation, it is useful to physically realize the transformation that, for a given PI input state $\sigma$, measures $j$, i.e., returns the angular momentum value $j$ with probability $p_j$, along with the corresponding post-measurement state $\sigma_j$.  In \cite{cervero2024memory}, this operation is called \emph{unitary} Schur sampling.

In practice, for many applications,  such as purification \cite{Cirac} and coherence distillation \cite{marvian2020coherence},  rather than returning the post-measurement state $\sigma_j$  as a subsystem with dimension $2j + 1$,  it is often more convenient and useful to have an output system whose Hilbert space is of constant size.  A natural and convenient way to achieve this is to encode the state $\sigma_j$ in the $(2j+1)$-dimensional totally symmetric subspace of $2j$ qubits.  The remaining $n-2j$ qubits (which is always an even number) can  
be prepared in pure singlet states, i.e., in the state $|\xi\rangle = (|01\rangle - |10\rangle)/\sqrt{2}$ (As we see below, this choice is made to respect the $\mathrm{SU}(2)$ symmetry). 
 That is, the goal is to realize the transformation
\be\label{Schur}
\mathcal{E}_{\text{Schur}}(\sigma)=\sum_{j=j_{\text{min}}}^{n/2} 
\widetilde{\sigma}_{j}\otimes 
\xi^{\otimes (\tfrac{n}{2}-j)}\otimes |j\rangle\langle j|_{\text{reg}}\ ,
\ee
where $\xi=|\xi\rangle\langle \xi|$,  $\widetilde{\sigma}_{j}$ is an unnormalized density operator in the totally symmetric subspace of $2j$ qubits satisfying $\Tr(\widetilde{\sigma}_{j})=p_j$, and we have appended the system with an additional (classical) register that keeps track of the value of $j$ as $|j\rangle_{\text{reg}}$.\footnote{Obviously, this register does not need to be a separate quantum device; it could be, for example, the controller of the experimental apparatus or simply the experimentalist's memory!} As we further explain below in Sec.\ref{Sec:compress}, if one is interested in obtaining state $\sigma_j$ encoded in $\lceil \log_2(2j+1) \rceil$ qubits, this can be achieved using an extra compression circuit, which converts $\widetilde{\sigma_j}/p_j $ to $\sigma_j$.

To explicitly write the matrix elements of $\widetilde{\sigma}_j$, we first consider a basis for the totally symmetric subspace of $2j$ qubits, namely, the states $|j, m\rangle$ for $m = -j, \dots, j$, the normalized eigenvectors of $\sum_{i=1}^{2j} \sigma_z^{(i)}/2$ in this subspace, also known as the Dicke states. In particular, $|j, m\rangle$ is the uniform superposition of all computational basis states in $\{|0\rangle, |1\rangle\}^{\otimes 2j}$ that contain $j + m$ qubits in state $|0\rangle$ and $j - m$ qubits in state $|1\rangle$ (see Appendix \ref{App:SU(2)} for the explicit formula). Then,
$$ |j, m\rangle\otimes |\xi\rangle^{\otimes (\tfrac{n}{2}-j)}\ \ , \ \ \ \ \ \text{for}\ \  \ m=-j,\cdots, j\ ,$$ 
is a basis for a single copy of an irrep of $\mathrm{SU}(2)$ with angular momentum $j$ in $\mathbb{C}^{2j+1}\otimes \mathbb{C}^{m(n,j)}$. More precisely, with respect to this decomposition, these vectors span the subspace $\mathbb{C}^{2j+1}\otimes |\eta_0\rangle$, where   $|\eta_0\rangle$ is a fixed state in $\mathbb{C}^{m(n,j)}$, the multiplicity subsystem associated to irrep $j$ (See Appendix \ref{App:SU(2)} for further discussion).

Then, according to Eq.(\ref{dec1}), the non-zero  matrix elements of  $\widetilde{\sigma}_j$ are determined by the matrix element of $\sigma$ via\footnote{It is worth noting that for state $\sigma=\rho^{\otimes n}$, the corresponding $\widetilde{\sigma}_j$,  up to a normalization, is independent of $n$, i.e., 
\be\nonumber
\frac{\langle j, m'|\widetilde{\sigma}_j|j, m\rangle}{m(n,j)} =  \left[\frac{1-\Tr(\rho^2)}{2}\right]^{\tfrac{n}{2}-j} \times \langle j, m'|\rho^{\otimes 2j}|j, m\rangle\ .
\ee
Therefore, while in general, the probability of observing angular momentum $j$ is determined by the number of given copies of state $\rho$ and its purity $\Tr(\rho^2)$, the corresponding reduced state in the sector with angular momentum $j$ is independent of $n$.} 
\be\label{matrix}
\frac{\langle j, m'|\widetilde{\sigma}_j|j, m\rangle}{m(n,j)}=\big[\langle j, m'|\langle\xi |^{\otimes (\tfrac{n}{2}-j)}\big] \sigma \big[|j, m\rangle |\xi\rangle^{\otimes (\tfrac{n}{2}-j)}\big]\ .
\ee

Eq.~(\ref{Schur}), together with Eq.~(\ref{matrix}), fully defines the action of the channel $\mathcal{E}_{\text{Schur}}$ on any PI state.   Explicitly, 
\be
\mathcal{E}_{\text{Schur}}(\sigma)=\sum_{j=j_{\text{min}}}^{n/2} 
M_j \sigma M_j \otimes |j\rangle\langle j|_{\text{reg}}\ ,
\ee
where 
\be\label{Kraus}
M_j=\sqrt{m(n,j)}\left(\sum^{j}_{m=-j} |j,m\rangle\langle j,m|\right)\otimes 
\xi^{\otimes (\tfrac{n}{2}-j)}\ .
\ee
\color{black} 
This definition can be extended to a general $n$-qubit state $\sigma$ by first applying the  
channel $\mathcal{P}$, which projects any input state to a PI state, and then applying the channel in Eq.~(\ref{Schur}) to the resulting symmetrized state.
 \begin{figure}[h]
\centering
\[
\Qcircuit @C=1em @R=1em {
\lstick{\ket{0}}      & \gate{H} & \ctrl{1}         & \gate{H} & \meter \\
   & \qw      & \multigate{1}{\text{SWAP}} & \qw      & \qw    \\
  & \qw      & \ghost{\text{SWAP}}         & \qw      & \qw
}
\]
\caption{The quantum circuit for implementing the SWAP test. In the fully coherent version of the protocol discussed in Sec.\ref{coherent} the final measurement is omitted. }
\label{fig:swap-test-box}
\end{figure}
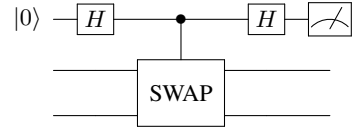

\section{Schur sampling via random SWAP tests}

Fig.~\ref{Fig} describes a protocol that can be implemented using only two-outcome two-qubit SWAP tests, which involve the projectors $\xi$ and $\mathbb{I} - \xi$ (See the circuit in Fig.\ref{fig:swap-test-box}). This test is often used to detect whether a given pair of qubits is in the same pure state or not. If they are, the outcome associated with $\mathbb{I} - \xi$, i.e., the projector onto the totally symmetric subspace, is always obtained. However, as we will see below, in order to explain and analyze this protocol, it is more convenient to think of this test as a way of detecting the singlet state $\xi$. Namely, at each step of the protocol, the goal is to detect a pair of qubits in the singlet state $|\xi\rangle$, then separate them from the rest of the qubits, and then continue the same steps with the remaining qubits. 

More precisely, as depicted in Fig.~\ref{Fig}, at each step of this protocol, we partition the $n$ qubits in the system into two subsets, $A_k$ and $B_k$, where $B_k$ contains $k$ pairs of detected and separated singlets and $A_k$ contains the remaining $n - 2k$ qubits. The classical register keeps track of the number of detected singlets. Specifically, when $k$ singlet pairs are detected, the register is in state $|n/2-k\rangle$, where 
$n/2-k$ denotes the current estimate of angular momentum $j$.

Initially, the register is in state $|n/2\rangle$, indicating that no singlet pairs have been detected yet (i.e., $k = 0$), which means that $A_0$ contains all $n$ qubits, while $B_0$ is empty. At each step, we uniformly select a random pair of qubits from $A_k$, perform a SWAP test, and if the pair is found to be in the singlet state, we move them to $B_k$ and update the register state as $|j\rangle \rightarrow |j - 1\rangle$.  The overall process defines a quantum channel  $\mathcal{E}_{\text{SWAP}}$. In particular, if the qubits in $A_k$ are in the state $\omega_{A_k}$, then after the next round of the protocol, the global state becomes
\begin{align}
&\mathcal{E}_{\text{SWAP}}(\omega_{A_k}\otimes \xi^{k}\otimes |\frac{n}{2}-k\rangle\langle \frac{n}{2}-k|)\\ &= \mathcal{F}_k(\omega_{A_k})\otimes \xi^{k}\otimes |\frac{n}{2}-k\rangle\langle \frac{n}{2}-k|\nonumber\\ &\ \  +\mathcal{S}_k(\omega_{A_k})\otimes \xi^{k+1}\otimes |\frac{n}{2}-k-1\rangle\langle \frac{n}{2}-k-1|\nonumber\ ,
\end{align}
where $\mathcal{F}_k$ and $\mathcal{S}_k$ are completely-positive quantum operations, which correspond to the cases where a singlet is not detected and is detected, respectively.\footnote{  In particular, for $r\neq s \le n-2k$, the Kraus operators of $\mathcal{F}_k$ and $\mathcal{S}_k$ are respectively $F_{k; r,s}=c_{n,k}(\mathbb{I}-|\xi\rangle\langle\xi|_{rs})$ and $S_{k; r,s}=c_{n,k} \langle\xi|_{rs}$,  where $c_{n,k}=\binom{n-2k}{2}^{-1/2}$ is a normalization factor and,   we have suppressed the identity operators on the rest of qubits.}  Then, after applying $T$ random SWAP tests on initial state $\tau$, we obtain the state
\begin{align}\label{def}
\mathcal{E}^T_{\text{SWAP}}&\left(\tau\otimes |\tfrac{n}{2}\rangle\langle \tfrac{n}{2}|\right) =  
\sum_{k=0}^{\lfloor\tfrac{n}{2}\rfloor} \omega_{A_{k}} \otimes \xi^{\otimes k} \otimes |\tfrac{n}{2}-k\rangle\langle \tfrac{n}{2}-k|\ ,
\end{align}
where  $\omega_{A_k}$ is the unnormalized state of $2j'=n-2k$ qubits in $A_{k}$, and $\Tr(\omega_{A_k})$ is the probability of detecting $k$ singlets.

As mentioned earlier, the state of the qubits in $A_k$ can be further compressed to $\lceil \log_2(2j+1) \rceil$ qubits, e.g., using circuits that prepare superpositions of Dicke states (see Sec.~\ref{Sec:compress}). However, it is important to emphasize that for many applications, such as qubit purification, this additional compression is irrelevant.

In the following, we show that in the limit $T\rightarrow \infty$, this channel converges to $\mathcal{E}_{\text{Schur}}$, and for $T\gg  n\ln n$, the error is exponentially small in $T$.\\

\section{Analyzing the Protocol}

\subsection{$\mathrm{SU}(2)$ Symmetry implies Reversibility}

This protocol is implemented using only a sequence of projective measurements, specifically, the SWAP test, which respects the $\mathrm{SU}(2)$ symmetry; that is, its projectors $\xi$ and $\mathbb{I} - \xi$ commute with $U \otimes U$ for any single-qubit unitary $U$. It is well known that under closed-system dynamics respecting $\mathrm{SU}(2)$ symmetry, the angular momentum operators $J_x$, $J_y$, and $J_z$ are conserved. A similar property holds for projective measurements that respect the symmetry: if the measurement outcome is discarded and we consider the average post-measurement state, the angular momentum operators remain conserved. This immediately implies that such projective measurements fully preserve PI information, because, according to Schur-Weyl duality, any PI state can be written as a linear combination of $U^{\otimes n}$ for $U \in \mathrm{SU}(2)$, or equivalently, as a polynomial in the angular momentum operators $J_x$, $J_y$, and $J_z$. In the following, we present this argument and its implications more formally. We note that a similar property is satisfied by the Schur transform, 
$\mathcal{E}_{\text{Schur}}$. Its Kraus operators $M_j \otimes |j\rangle$ commute with $U^{\otimes n}$, a property sometimes referred to as the strong symmetry of the channel, which in turn implies that it preserves the angular momentum operators (See Appendix \ref{App:SU(2)} for further discussion).

Consider an arbitrary $n$-qubit PI state restricted to a single angular momentum sector $j$, i.e.,
\be\label{input}
\tau_j=\sigma_{j}\otimes\frac {\mathbb{I}_{m(n,j)}}{m(n,j)}=\mathcal{P}\big(\widetilde{\sigma}_j\otimes \xi^{\otimes (\tfrac{n}{2}-j)}\big)\ .
\ee
Here, the first tensor product is with respect to the component $\mathbb{C}^{2j+1} \otimes \mathbb{C}^{m(n,j)}$ in the decomposition of Eq. (\ref{dec}), and $\widetilde{\sigma}_j$ is the state associated with $\sigma_j$ via Eq.  (\ref{matrix}) (namely, the encoding of $\sigma_j$ in the totally symmetric subspace of $2j$ qubits).

Thanks to the aforementioned $\mathrm{SU}(2)$ symmetry of the measurement projectors, the support of the output 
$\mathcal{E}^T_{\text{SWAP}}(\tau_j)$
is also restricted to the  subspace with angular momentum $j$, i.e., $\mathbb{C}^{2j+1} \otimes \mathbb{C}^{m(n,j)}$. Furthermore, since the measurement projectors act trivially on $\mathbb{C}^{2j+1}$, for each possible outcome of the SWAP tests, the post-measurement state is still of the form $\sigma_j \otimes \beta_k$ with respect to this decomposition, where $\beta_k$ depends on the outcomes. In other words, the reduced state in each irrep of $\mathrm{SU}(2)$ remains fully protected.

Now, in terms of $\omega_{A_k}$, the unnormalized state of qubits in $A_k$ as defined in Eq.~(\ref{def}), this means that this state is also restricted to the sector with angular momentum $j$ of the $n - 2k$ qubits in $A_k$ (recall that the singlet is invariant under $\mathrm{SU}(2)$ and does not change the angular momentum of the system).  
In other words, the original input state $\tau_j$ can be fully recovered from the remaining qubits in $A_k$ simply by appending $k$ singlets and applying a random permutation, as
\begin{align}\label{mixed}
 \mathcal{D}\left(\frac{\omega_{A_{k}}}{\Tr(\omega_{A_{k}})}\right):=  \mathcal{P}\left(\frac{\omega_{A_{k}}}{\Tr(\omega_{A_{k}})} \otimes \xi^{\otimes k}\right)=\tau_j\ ,
 \end{align}
which holds for all  $k$ with non-zero probability. This can be seen by noting that, under random permutations, the state of the subsystem $\mathbb{C}^{m(n,j)}$ is mapped back to the maximally mixed state.  

For a general PI state $\tau$ that is not restricted to a single angular momentum sector $j$, a weaker notion of reversibility holds. Specifically, if we discard the register that records $k$, the number of detected singlets, and then apply a random permutation to the $n$ qubits, we recover the original PI state $\tau$, i.e.,
\be\label{rev4}
\mathcal{P}\circ\Tr_{\text{reg}}\circ\mathcal{E}^T_{\text{SWAP}}(\tau)=\tau\ ,
\ee
which follows from the $\mathrm{SU}(2)$ symmetry of the SWAP test (See Appendix \ref{App:SU(2)}). \color{black}

Finally, it is worth noting that the map $\mathcal{D}$ defined by Eq.~(\ref{mixed}) is also reversible on any PI input state. Indeed, for an arbitrary PI state $\alpha$ of $n-2k$ qubits, the input state $\alpha$ can be recovered from $\mathcal{D}(\alpha)$ simply by applying the Schur transform and discarding the register and the $k$ singlets, as
\be\label{rev}
\Tr_{k \text{ singlets}} \circ \Tr_{\text{reg}} \circ \mathcal{E}_{\text{Schur}} \circ \mathcal{D}(\alpha) = \alpha\ .
\ee
See Appendix \ref{App:SU(2)} for further discussions. We emphasize that the above arguments rely solely on $\mathrm{SU}(2)$ symmetry and Schur-Weyl duality, and therefore admit broad generalizations. In particular, the reversibility holds for any sequence of SWAP tests on qudits
 (see Appendix~\ref{App:SU(2)}).

\subsection{Proof of Convergence }

Remarkably, these simple observations allow us to fully analyze this protocol.   Suppose we start with the initial state $\tau_j$ in Eq.~(\ref{input}) and detect $k$ singlets, which means the register is in state $|j' = n/2 - k\rangle$, and $A_k$ contains $n-2k=2j'$ qubits. 
 Then,
\begin{itemize}
    \item $k>n/2-j$: This will never happen, i.e., its probability is  $\Tr(\omega_{A_k})=0$, because otherwise the number of qubits in $A_k$ would be less than $2j$, and they cannot have angular momentum $j$.
    
    \item $k<n/2-j$: State $\omega_{A_k}$ is orthogonal to the totally symmetric subspace of the $n - 2k$ qubits in $A_k$, as they correspond to different total angular momenta, specifically $j \ne j' = n/2 - k$. This implies that a random SWAP test on $A_k$ has a nonzero probability of detecting a singlet, which will then be moved to the $B$ part, thereby increasing $k$ by 1. Using the permutational and $\mathrm{SU}(2)$ symmetries of the protocol, we show that this probability is a simple function of $j$ and $j'$, denoted by $e_{j'}(j)$ in Eq.(\ref{prob}). In particular, it is independent of $T$, the state of the qubits, and other parameters.

    \item $k = n/2 - j$: Since state  $\omega_{A_k}$ of $A_k$ is restricted to the totally symmetric subspace of the  $2j$ qubits in $A_k$, the SWAP tests fail to detect any additional singlet. Furthermore, the normalized state is  $\omega_{A_k}/\Tr(\omega_{A_k}) = \widetilde{\sigma}_j$.
\end{itemize}

The last claim follows immediately from the fact that the subsystem $\mathbb{C}^{2j+1}$ in the decomposition of Eq.~\eqref{dec} remains completely unaffected by the protocol.
  Equivalently, this follows from the reversibility statement in Eq.~(\ref{rev}), which implies that for any PI input state, the output of the channel $\mathcal{D}$ uniquely determines the input. Then, comparing Eq.~(\ref{mixed}) and Eq.~(\ref{input}) shows that, when $k = n/2 - j$, we have $\omega_{A_k} / \Tr(\omega_{A_k}) = \widetilde{\sigma}_j$.

\begin{figure}
    \centering
    \includegraphics[width=1\linewidth]{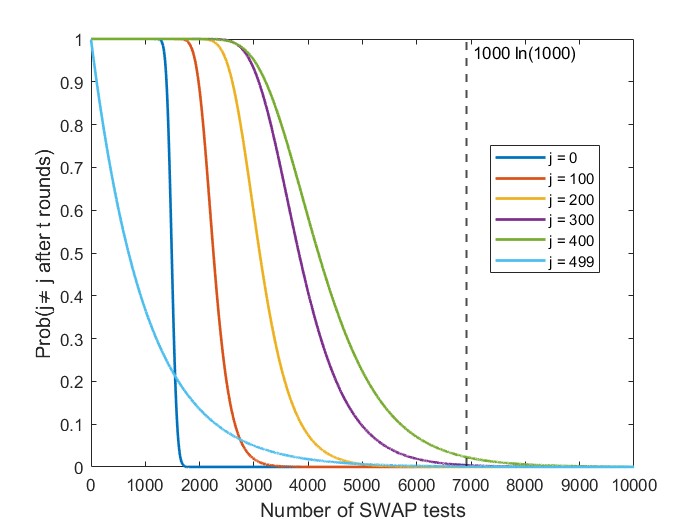}
\includegraphics[width=1\linewidth]{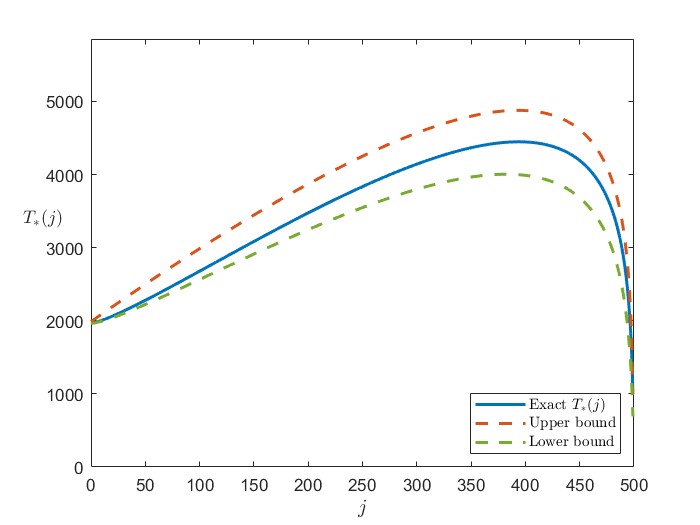}
 \caption{Eq.~(\ref{TD}) shows that the error in this protocol, as quantified by the trace distance with the output of the Schur transform, for input states restricted to a sector with angular momentum $j$, is exactly equal to the probability that not all of the existing $\frac{n}{2} - j$ singlets in the system have been detected.   
\textbf{Top} plot shows this probability as a function of $T$, the number of SWAP tests, for a system with $n=1000$ qubits. 
\textbf{Bottom} plot shows $T_\ast(j)$, the expected number of SWAP tests required to detect all ${n}/{2} -j$ singlets, again for the same number of $n=1000$ qubits. }
    \label{fig:probj'neqj}
\end{figure}

In summary, we conclude that (i) if all $n/2 - j$ singlets in the system have been detected, the remaining qubits are in the desired state $\widetilde{\sigma}_j$, and (ii) in the limit $T \rightarrow \infty$, all $n/2 - j$ singlets will eventually be detected, thereby yielding  $\widetilde{\sigma}_j$.  

Furthermore, remarkably, we find that the performance of this protocol is fully determined by the distribution of the number of detected singlets. 
 In particular, the trace distance between the output of this protocol for the input state $\tau_j$ and the desired output $\mathcal{E}_{\text{Schur}}(\tau_j)$ is equal to the total variation distance between this distribution and the delta distribution concentrated at $j$, which is equal to the probability that the register is not in the correct state $|j\rangle$, i.e.,

\be\label{TD}
D\big(\mathcal{E}_{\text{Schur}}(\tau_j)\ ,  \mathcal{E}^T_{\text{SWAP}}\left(\tau_j \otimes \left|\tfrac{n}{2}\right\rangle\left\langle \tfrac{n}{2}\right|\right)\big)=\Pr(j'\neq j; T)  \ ,
\ee
where $D(\gamma_1,\gamma_2)=\tfrac{1}{2}\|\gamma_1-\gamma_2\|_1$ is the trace distance, and $\Pr(j'\neq j; T) =1-\Tr(\omega_{A_{n - 2j}})$. Furthermore, according to Eq.(\ref{dec1}), a general PI state $\sigma$ can be decomposed as a convex combination $\sigma=\sum_j p_j \tau_j$ of states $\tau_j$ that are restricted to sectors with a fixed angular momentum. Then, the triangle inequality implies the trace distance of the output states $\mathcal{E}_{\text{Schur}}(\sigma)$ and $\mathcal{E}^T_{\text{SWAP}}\left(\tau\otimes \left|\tfrac{n}{2}\right\rangle\left\langle \tfrac{n}{2}\right|\right)$,  is bounded by $\sum_{j} p_j\times \Pr(j'\neq j; T)$. 
As we prove more formally below, $\Pr(j'\neq j; T)$ vanishes in the limit $T$ goes to infinity, which in turn implies that for any $n$-qubit PI state $\sigma$, the output of  $\mathcal{E}^T_{\text{SWAP}}$ 
converges to the output of the ideal Schur transform, i.e.,
\be
\lim_{T\rightarrow \infty} \mathcal{E}^T_{\text{SWAP}}\left(\sigma\otimes |\tfrac{n}{2}\rangle\langle \tfrac{n}{2}|\right)=\mathcal{E}_{\text{Schur}}(\sigma)\ .
\ee
Indeed, as we explain in more detail in the following section and expand further in Appendix \ref{Appendix:D}, the trace distance between the two sides decreases monotonically and exponentially with $T$ (See Eq.(\ref{final-bound})).

\section{Convergence Rate and Error Analysis}
As we showed in the previous section and highlighted in Eq.~(\ref{TD}), the error in this protocol is fully determined by the probability of finding all the singlets in the input state. In the following, we analyze this probability and the number of SWAP tests needed to make it arbitrarily small.

\subsection{Probability of Detecting a Singlet}
First, we calculate the conditional probability that, for the initial state $\tau_j$, the SWAP test $T+1$ detects the $(k+1)$'th singlet, given that $k$ singlets have been detected before. Since the SWAP test respects $\mathrm{SU}(2)$ symmetry, and each SWAP test is performed on a pair of qubits randomly and uniformly chosen from the $2j'=n - 2k$ qubits in $A_k$, the outcome probability does not change if, prior to the measurement, we apply any random permutation of these qubits or any global rotation $U^{\otimes 2j'}$ for arbitrary $U \in \mathrm{SU}(2)$. As we further explain in Appendix~\ref{Appendix:probabilities}, it follows that this probability is equal to the probability of detecting a singlet by performing a SWAP test on the state that is maximally mixed in the sector with angular momentum $j$ of $2j'$ qubits, which is equal to

\be\label{cond}
\Pr(k+1 ; T+1 \mid k ; T)=\frac{ \Tr(\omega_{A_{k}}E_{j'})}{\Tr(\omega_{A_{k}})}=e_{j'}(j) \ ,
\ee
where 
\begin{align}\label{prob}
e_{j'}(j)=\frac{j'(j'+1)-j(j+1)}{2j'(2j'-1)}> 0\ ,
\end{align}
is the eigenvalue of 
 the average of projectors onto the singlets, i.e., the operator
\be
E_{j'} ={\binom{2j'}{2}}^{-1} 
\sum_{r<s} \xi_{r,s}\ ,
\ee
in the sector with angular momentum $j$,  where the summation is  over $r, s$ in the range $1 \le r < s \le 2j'$.  Therefore, as expected, this probability is positive for $j'>j$, or equivalently, $k<n/2-j$. Moreover, when the test fails to detect a singlet, this conditional probability remains unchanged, i.e., it is independent of $T$.\\

\begin{figure}
    \centering
    \includegraphics[width=1\linewidth]{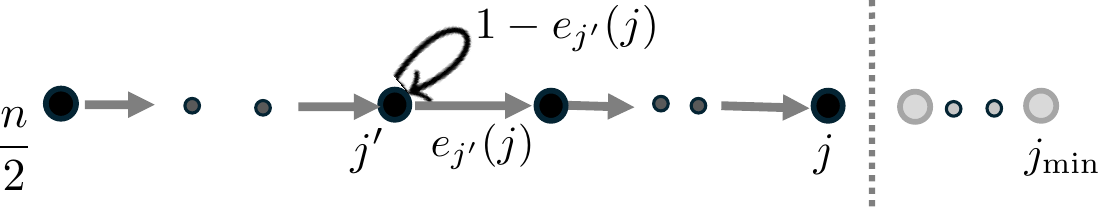}
\caption{The nodes represent the state of register $|j'\rangle$, or, equivalently,   the number of detected singlets, namely $k=n/2-j'$  singlets. Initially, the register is at $j'=n/2$, which means we have not detected any singlet yet.  If the input state $\tau_j$ is restricted to the sector with angular momentum $j$, then in the limit $T\gg 1 $ SWAP tests, the register goes to state $|j\rangle$ with probability 1, i.e., we detect all the existing $n/2-j$ singlets. The process can be understood as a Markov chain: When the register is in state $j'$ the probability that a SWAP test detect a singlet is $e_{j'}(j)$ given in Eq.(\ref{prob}), which means with probability $1-e_{j'}(j)$ the register remains in the same state.} 
    \label{fig:enter-label}
\end{figure}

\subsection{Sharp Transition in the Probability of Error}
Let $X_r: r=1,\cdots, k$ be the random variable corresponding to the number of SWAP tests needed to detect the $r$'th singlet.  Each SWAP test detects $r$'th singlet with probability $e_{j'}(j)$, with $j'=n/2-r+1$, 
which defines a geometric random variable with the mean 
\be\label{mean1}
\langle X_r\rangle=e_{j'}(j) \sum_{t=1}^{\infty} t [1-e_{j'}(j)]^t  = \frac{1}{e_{j'}(j)}\ .
\ee
Hence, the mean number of the total SWAP tests needed to detect all $k=n/2-j\ge 1$ singlets is
\be\label{eqn:15}
T_{\ast}(j)=\sum_{j'>j}^{n/2} \frac{1}{e_{j'}(j)}\le 2(n-j)+(2j-2)  \ln(\frac{n}{2}-j)\ ,
\ee
where this bound is shown in Appendix~\ref{Appendix C}, and is proven to be tight (See the bottom plot in Fig. \ref{fig:probj'neqj}).

As the example in Fig.\ref{fig:probj'neqj} shows there is a sharp transition around this mean value: For $T\ll T_{\ast}(j)$, the probability of error $\Pr(j'\neq j; T)$ is approximately one, and for $T\gg T_{\ast}(j)$ this probability is approximately zero.    
To analyze this, we note that
\be \label{proboferror}
\text{Pr}(j'\neq j; T)=\Pr(X_1 + \cdots + X_k>T)\ .
\ee 
Here, $X_1 + \cdots + X_k$  is the random variable corresponding to the total number of SWAP tests needed for finding exactly $k=n/2-j$ singlets. Since this random variable is the sum of a finite number of independent geometric random variables with finite means and variances, the probability that it takes values larger than its mean $T_\ast(j)$, decays exponentially. Using standard techniques, namely, applying  Chernoff's bound for the sum of geometric random variables \cite{jansontail}, in Appendix \ref{Appendix:D}, we study this exponential decay.

Finally, putting everything together in Appendix~\ref{Appendix:D},  we find that for a general PI state $\sigma$, and $T \geq n \ln n$, it holds that

\begin{align}\label{final-bound}
D\big(\mathcal{E}_{\text{Schur}}(\sigma),\ \mathcal{E}^T_{\text{SWAP}}\left(\sigma \otimes \left|\tfrac{n}{2}\rangle\langle \tfrac{n}{2}\right|\right)\big) \le n\exp\left(-\frac{T}{2n}\right)\ ,
\end{align}
\color{black}
which implies Eq.(\ref{T-bound}). In conclusion, once the number of SWAP tests exceeds $n \ln n$, the error converges to zero exponentially fast in $T$.

\section{Application: Qubit  Purification}\label{Sec:App:Pur}
Suppose we apply this protocol to the initial state $\sigma = \rho^{\otimes n}$, where 
$\rho = (1 - p)|\psi\rangle\langle\psi| + p \mathbb{I}/2$, 
with $|\psi\rangle$ an unknown pure state and $0 < p < 1$. Suppose that, after performing $T$ SWAP tests and discarding the detected singlets, we select one of the remaining qubits uniformly at random. Then, as we show in Appendix~\ref{App:Fid}, the fidelity of the resulting output state with the target pure state $|\psi\rangle$ is
\begin{align}\label{fid}
f(n,T)&=\frac12 + \sum_{j} p_j  (f_j-\frac12)  \sum_{j'\ge j}  \Pr(j'|j,T)\times \frac{j}{j'} \ . 
\end{align}
Here, $p_j = \Tr(\rho^{\otimes n} \Pi_j)$ is the probability that $\rho^{\otimes n}$ is found in the sector with total angular momentum $j$, and $f_j$ denotes the optimal achievable fidelity with the target state $|\psi\rangle$, conditioned on the system being found in this sector. These functions, previously computed by Cirac \textit{et al.}~\cite{Cirac}, are presented in Appendix~\ref{App:Fid}. Furthermore, $\Pr(j'|j,T)$ denotes the probability that, for an initial state restricted to the subspace with total angular momentum $j$, that is, a state of the form $\tau_j$ in Eq.~(\ref{input}), exactly $n/2 -j'$ singlets are detected after performing $T$ SWAP tests, implying that the estimated angular momentum is $j'$.

As argued earlier, in the limit $T \rightarrow \infty$, the probability $\Pr(j'|j,T)$ converges to the Kronecker delta $\delta_{j,j'}$. In this case, we recover the optimal fidelity previously obtained in~\cite{Cirac}, which is achievable via the Schur transform, namely,
\be
f_{\text{opt}}(n)= \sum_j p_j f_j=1 - \frac{1}{2n} \frac{p}{(1 - p)^2} + \mathcal{O}\left(\frac{1}{n^2}\right)\ .
\ee
From the form of Eq.~(\ref{fid}), it is easy to see that unless the input state $\rho$ is already pure, the fidelity $f(n,T)$ is strictly less than the optimal fidelity $f_{\text{opt}}(n)$ for any finite $T$. However, Eq.~(\ref{final-bound}) immediately implies that the difference between them is exponentially small in $T$, that is,

\be \label{fideff}
f_{\text{opt}}(n) - f(n,T) \le n \exp\left(-\frac{T}{2n}\right)\ .
\ee

See Fig.~\ref{fig:fidconvergence} for an example with a system of $n = 100$ qubits. \\

We briefly compare the performance of our protocol with that of Childs \textit{et al.}~\cite{Childs2025streamingquantum}, which, as discussed in the introduction, also relies on SWAP tests. However, unlike our approach, their protocol discards one of the two qubits whenever the pair is projected onto the triplet subspace. Assuming access to copies of a noisy qubit state $\rho = (1-p)|\psi\rangle\langle\psi| + p\,\mathbb{I}/2$, Ref.~\cite{Childs2025streamingquantum} analyzes the expected number of copies of $\rho$ required to prepare a state with fidelity at least $1 - \epsilon$ with $|\psi\rangle$, and shows that this number is upper bounded by
\[
n \le 3630\, \epsilon^{-1} (1 - p)^{-8\ln 2} \, ,
\]
where $8 \ln 2 \approx 5.54518$. Note that in their protocol, the number of SWAP tests, which determines the gate complexity, is on the order of the number of copies used.

In contrast, since our protocol converges to the ideal Schur transform in the limit $T \rightarrow \infty$, Eq.~(\ref{opt-Fid}) from Ref.~\cite{Cirac} implies that we can achieve the same error $\epsilon \ll 1$ using roughly
\[
n_{\epsilon} = \frac{1}{2} \epsilon^{-1} \, p (1 - p)^{-2}
\]
copies of $\rho$, assuming there is no restriction on the number of SWAP tests. If we limit the number of SWAP tests, the same fidelity can be achieved using roughly twice this number of copies, with a total of $\lceil 2n_{\epsilon} \ln{\frac{2n_{\epsilon}}{\epsilon}} \rceil$ SWAP tests.

Crucially, in the regime $p \rightarrow 1$, i.e., when the input noise is strong, the expected sample and gate complexities of the protocol in Ref.~\cite{Childs2025streamingquantum} scale as $(1 - p)^{-5.54518}$, while in our protocol, both complexities scale as $(1 - p)^{-2}$, up to lower-order terms.

\section{Discussion}

Remarkably, we saw that just by performing a simple, well-known primitive measurement, namely, the SWAP test, it is possible to realize the Schur transform on PI states and achieve the optimal qubit purification. Furthermore, the number of required SWAP tests to accomplish this, up to a possible logarithmic improvement, matches the minimum number of operations needed by any protocol interacting with $n$ qubits. In particular, an error $\epsilon$ in the output, e.g., as quantified in terms of the trace distance, can be achieved using only 
$
T = 2 n \ln \left( n \epsilon^{-1} \right)$ SWAP tests. 

We discussed the application of this protocol in the context of qubit purification and showed how the optimal purification can be achieved with this protocol. In addition to purification, another important application of this protocol is in the context of tomography \cite{hu2024sample}. Namely, without losing any information, one can run this protocol to discard qubits and continue with the remaining purer qubits.

While in this work we focused on the simplest version of this algorithm, there are other interesting variants. In the following, we discuss two such variants of this protocol: a coherent version of the protocol, and a version which further compresses the output states into $\lceil\log_2 (n + 1)\rceil$ qubits. In future work, we aim to further analyze other variants of this protocol, specifically:  
(i) a derandomized version of the algorithm, where the SWAP tests are selected deterministically, which could be useful to reduce the complexity and to simplify the coherent version discussed below; (ii) protocols that can be realized with random SWAP tests on qudits.

\subsection{Compression}\label{Sec:compress}

As we saw before, this protocol returns a classical register that records the number of detected singlets $k$, or equivalently, the angular momentum $j = n/2 - k$. Additionally, it returns $2j = n - 2k$ qubits in a state that, with high probability, is confined to the $(2j + 1)$-dimensional totally symmetric subspace. For some applications, such as purification, this is precisely the required output format. In principle, however, for other applications, it may be useful to compress the state of the totally symmetric subspace of the output qubits into $\lceil \log_2(2j + 1) \rceil$ qubits. This can be achieved, for example, using circuits for preparing superposition of Dicke states~\cite{liu2024low}. In particular, to do so, it suffices to first implement the sorting transformation, which maps an arbitrary state in the symmetric subspace of $2j$ qubits as
\begin{align}
\sum_{m=-j}^j\psi_m |j,m\rangle &\longrightarrow \sum_{m=-j}^j\psi_m |0\rangle^{\otimes (j+m)}\otimes |1\rangle^{\otimes (j-m)}  \ ,
\end{align}
Then, the output state, which has a unary encoding,  can be easily converted to a binary encoding. The entire conversion can be achieved using $\mathcal{O}(j\log j)$ gates, e.g., by applying the method of \cite{liu2024low}. 
In conclusion, by composing these conversion algorithms with our random SWAP protocol, we can implement Schur transform on PI states, with $\mathcal{O}(n \log n)$ gates.  

\subsection{A Fully Coherent Protocol}\label{coherent}
So far, we have focused on PI input states. For general states, the above protocol erases information that is not invariant under permutations. More precisely, with respect to the decomposition in Eq. (\ref{dec}), it dephases the state across different $j$ sectors $\mathbb{C}^{2j+1} \otimes \mathbb{C}^{m(n,j)}$, and furthermore, erases information encoded in the subsystems $\mathbb{C}^{m(n,j)}$ corresponding to the irreducible representations of $\mathbb{S}_n$. In contrast, the Schur transform~\cite{weakchilds, SchurTransformHarrow, harrowthesis} fully preserves this information.

Interestingly, as we discuss in Appendix~\ref{App:Coh},  the above protocol can also be implemented in a fully coherent manner, without any measurements. In this case, similar to the Schur transform, the entire process is unitary and therefore fully reversible. In particular, in the coherent version, the outcome of each SWAP test is not measured but instead stored in an ancilla qubit. Furthermore, the pair of qubits undergoing the SWAP test is also selected in a coherent fashion. Then, controlled unitaries are employed to determine the next steps based on the previous ones, as we further discuss in Appendix~\ref{App:Coh}.

\section*{Acknowledgments}

We acknowledge helpful discussions with other members of our group, especially Sujay Kazi and Shiv Akshar Yadavalli.  We also acknowledge useful comments by Zhaoyi Li.  We also acknowledge support from  Army Research Office
(W911NF-21-1-0005), NSF Phy-2046195, NSF FET-2106448, and 
NSF QLCI grant OMA-2120757. SB thanks Justin Htay and Hirad Tabatabaei for proofreading the manuscript.

\bibliographystyle{unsrt}
\bibliography{Bib}

\newpage

\appendix
\onecolumngrid
\color{black}

\newpage
\section{From Symmetry to Reversibility}\label{App:SU(2)}

\subsection{Review of Useful Facts about the Representation Theory of $\mathrm{SU}(2)$ and $\mathbb{S}_n$
}

Here, we review some useful facts about the representation theory of $\mathrm{SU}(2)$  and $\mathbb{S}_n$.\\

\begin{enumerate}

\item A convenient way to understand the spin-$j$ representation of $\mathrm{SU}(2)$ is via the totally symmetric subspace of $(\mathbb{C}^2)^{\otimes 2j}$, which has dimension $2j+1$. The representation $U \mapsto U^{\otimes 2j}$ of $\mathrm{SU}(2)$ acts irreducibly on this subspace, realizing the spin-$j$ representation. A natural basis for this subspace is given by the eigenbasis of the operator $J_z$. The eigenvalues of $J_z$ are non-degenerate and take the values $m = -j, -j+1, \ldots, j$, with the corresponding eigenvector denoted by $\ket{j, m}$.
\be\label{Dicke}
|j,m\rangle = \frac{1}{\sqrt{\binom{2j}{j+m}}} \sum_{\substack{b_1,\cdots, b_{2j} \in \{0,1\} \\ \sum_{k=1}^{2j} b_k = j-m}} |b_1\rangle\cdots |b_{2j}\rangle\ ,
\ee
which is the uniform sum of all bit strings $b_1\cdots b_{2j}$ with Hamming weight $j-m$. These states are also called  Dicke states. \\

\item When two irreducible representations of $\mathrm{SU}(2)$ with angular momenta $j_1$ and $j_2$ are combined, the resulting decomposition contains all total angular momenta $j = |j_1 - j_2|, |j_1 - j_2| + 1, \ldots, j_1 + j_2$, each appearing with multiplicity one.
That is
 \be
\mathbb{C}^{2j_1+1}\otimes \mathbb{C}^{2j_2+1} \cong \bigoplus_{j=|j_1-j_2|}^{j_1+j_2}\mathbb{C}^{2j+1} \ . 
 \ee
 In particular, because $(U\otimes U)|\xi\rangle=|\xi\rangle$ for all $U \in \mathrm{SU}(2)$, adding singlets which have angular momentum $j=0$, does not change the angular momentum of the system. \\

\item The fact that $|j, m\rangle \otimes |\xi\rangle^{\otimes (n - 2j)}$ is not entangled with respect to $\mathbb{C}^{2j+1} \otimes \mathbb{C}^{m(n, j)}$ can be seen, for example, by noting that within each irrep of SU(2), the eigenvalues of $J_z$ are non-degenerate. Moreover, the fact that the associated vector in $\mathbb{C}^{m(n, j)}$ is identical for all elements of this basis follows from the observation that these basis elements can be transformed into one another using the PI ladder operator $J_-=\sum_{i=1}^n |1\rangle \langle 0|_i/2$, which act as the identity operator on $\mathbb{C}^{m(n, j)}$.\\

\item Recall that the SWAP operator on a pair of $d$-dimensional system is given by
\be
\text{SWAP}=\sum_{i,j=0}^{d-1} |i\rangle\langle j|\otimes |j\rangle\langle i|\ .
\ee
SWAP unitaries on $n$ qudits generate a representation of the symmetric group $\mathbb{S}_n$, which will be denoted as $\textbf{P}(\pi): \pi\in\mathbb{S}_n$. In particular, the SWAP operation on qubits $r$ and $s$ implements the transposition $(rs)$.  \\

\item Any input state $\tau$ of $n$ qubits can be projected onto its permutationally invariant version by applying a random permutation chosen uniformly at random from the group $\mathbb{S}_n$. More precisely, this is achieved by the twirling map
\be
\mathcal{P}(\tau) = \frac{1}{n!} \sum_{\pi \in \mathbb{S}_n} \textbf{P}(\pi) \, \tau \, \textbf{P}(\pi)^\dagger,\ \ 
\ee
where the summation is over all elements of $\mathbb{S}_n$. Then, $\mathcal{P}(\tau)$ is a PI state, which with respect to decomposition in Eq.(\ref{dec}) takes the form
\be
\mathcal{P}(\tau)=\bigoplus_{j=j_\text{min}}^{n/2} p_j \big(\tau_j\otimes \frac{\mathbb{I}_{m(n,j)}}{m(n,j)} \big)\ ,
\ee
where $p_j=\Tr(\Pi_j \tau)$, and
\be
\tau_j=\frac{1}{p_j}\Tr_{\mathbb{C}^{m(n,j)}}(\Pi_j \tau \Pi_j) \ ,
\ee
and the partial trace is with respect to the decomposition $\mathbb{C}^{2j+1}\otimes \mathbb{C}^{m(n,j)}$ on the subsystem  $\mathbb{C}^{m(n,j)}$.\\

\item According to the Schur-Weyl duality, the set ${U^{\otimes n} : U \in \mathrm{SU}(2)}$ spans the space of all PI operators on $(\mathbb{C}^2)^{\otimes n}$, which in turn implies any PI operator $B$ is uniquely specified by its Hilbert-Schmidt inner product with this set, i.e., the characteristic function $\Tr(B U^{\otimes n}): U \in \mathrm{SU}(2)$.   

\end{enumerate}

\subsection{ $\mathrm{SU}(2)$ symmetry implies reversibility (Proofs of Eqs.(\ref{rev4}, \ref{rev}))}

The fact that the Kraus operators of both the Schur transform $\mathcal{E}_{\text{Schur}}$ and   $\mathcal{E}_{\text{SWAP}}$ commute with $U^{\otimes n}: U\in \mathrm{SU}(2)$ implies that for any operator $B$,
\be\label{commute}
U^{\otimes n}\mathcal{E}(B)=\mathcal{E}(U^{\otimes n} B)  \ \  \ \ \  \text{and}\  \ \  \ \ \mathcal{E}(B)U^{\otimes n}=\mathcal{E}(B U^{\otimes n})\ ,
\ee
where we have suppressed the subscripts. This property, which is sometimes referred to as the strong symmetry of the channel in the open systems literature, is also satisfied by the uniform twirling channel $\mathcal{P}$.

It follows that 
$$
\Tr\left(U^{\otimes n}\mathcal{P}\circ\Tr_{\text{reg}}\circ\mathcal{E}^T_{\text{SWAP}}(\tau)\right)=\Tr(U^{\otimes n}\tau)\ ,
$$
which, in turn, implies that for any PI state $\tau$, $\mathcal{P}\circ\Tr_{\text{reg}}\circ\mathcal{E}^T_{\text{SWAP}}(\tau)=\tau$, which proves  Eq.(\ref{rev4}).

Similarly, consider the channel
\be
\mathcal{D}_{m\rightarrow n}(\sigma)=\mathcal{P}(\sigma\otimes \xi^{\otimes (n-m)/2})\ ,
\ee
which maps $m$-qubit state $\sigma$ to $n>m$ qubit-state $\mathcal{D}_{m\rightarrow n}(\sigma)$. Then, for any PI $m$-qubit state $\sigma$,  
\be
\Tr(U^{\otimes n}\mathcal{D}_{m\rightarrow n}(\sigma))=\Tr(U^{\otimes n}\mathcal{P}(\sigma\otimes \xi^{\otimes (n-m)/2}))=\Tr(\sigma U^{\otimes m})\ ,
\ee
which means   $\mathcal{D}_{m\rightarrow n}(\sigma)$  uniquely specifies any PI $\sigma$. Indeed, According to Eq.(\ref{rev}) state $\sigma$ can be recovered from the output of channel $\mathcal{D}_{m\rightarrow n}(\sigma)$, using the Schur transform, namely 
\be\nonumber
\Tr_{\tfrac{n-m}{2} \text{ singlets}} \circ \Tr_{\text{reg}} \circ \mathcal{E}_{\text{Schur}} \circ \mathcal{D}_{m\rightarrow n}(\sigma) = \sigma\ .
\ee
To see this, we again consider the Hilbert-Schmidt inner products of both sides with $U^{\otimes m}$, and use Eq.(\ref{commute}) for $\mathcal{E}_{\text{Schur}}$ and the fact that singlets are invariant under $\mathrm{SU}(2)$.\\

\subsection{Reversibility after $\mathrm{SU}(d)$-invariant measurements}

Using Schur-Weyl duality for the representation of $\mathrm{SU}(d)$ on $(\mathbb{C}^d)^{\otimes n}$, the above results can be readily generalized to a system of $n$ qudits. In particular, for PI initial states, reversibility holds for any number of SWAP tests, regardless of which qudits are measured or what the measurement outcomes are.

Here, we state another useful variant of the above reversibility statements, which holds for qudits.

\begin{lemma}\label{lem:SU}
   Consider a quantum measurement defined by a set of measurement operators $\{M_r\}$ satisfying the completeness relation $\sum_r M_r^\dag M_r = \mathbb{I}$. Suppose each measurement operator is $\mathrm{SU}(d)$-invariant, i.e., $[M_r, U^{\otimes n}] = 0$ for all $U \in \mathrm{SU}(d)$. 
   \begin{enumerate}
       \item After performing this measurement on any PI state $\sigma$, we can recover the original state simply by forgetting the outcome, and applying a random permutation on qusits, i.e.
       \be
\mathcal{P}(\sum_{r} M_r \sigma M_r^\dag)=\sigma\ .
       \ee
       \item In the special case where the PI state is restricted to an isotypic component of an irrep of $\mathrm{SU}(d)$, or, equivalently,  $\mathbb{S}_n$ (e.g., the angular momentum sector $j$, $\mathbb{C}^{2j+1}\otimes \mathbb{C}^{m(n,j)}$ in the case of $\mathrm{SU}(2)$), the original state can be recovered from the post-measurement state $M_r \sigma M_r^\dag / \Tr(M_r \sigma M_r^\dag)$ for any outcome $r$ with non-zero probability. Furthermore, for any such post-measurement state and $U \in \mathrm{SU}(d)$,
    \be
       \Tr\left(U^{\otimes n} \sigma\right) = \Tr\left(U^{\otimes n} \frac{M_r \sigma M_r^\dag}{\Tr(M_r^\dag M_r \sigma)}\right) \ .
    \ee
   \end{enumerate}
\end{lemma}
In the case of $\mathrm{SU}(2)$, this lemma implies that if the initial state $\sigma$ is permutationally-invariant and lies entirely within an eigensubspace of the total angular momentum operator $J^2$, then, for any measurement outcome $r$ with non-zero probability, the expectation value of the angular momentum along any direction $\hat{n}$ is conserved,
\be
\Tr\left(J_{\hat{n}} \sigma\right) = \Tr\left(J_{\hat{n}} \frac{M_r \sigma M_r^\dag}{\Tr(M_r^\dag M_r \sigma)}\right) \ .
\ee
\begin{proof}
According to Schur-Weyl duality,  $U^{\otimes n}: U\in\mathrm{SU}(d)$ spans the space of PI operators. 

State $\mathcal{P}(\sum_{r} M_r \sigma M_r^\dag)$ is PI. Then, the fact that for $U\in\mathrm{SU}(d)$, 
\be
\Tr(U^{\otimes n}\mathcal{P}(\sum_{r} M_r \sigma M_r^\dag))=\Tr(U^{\otimes n}\sum_{r} M_r \sigma M_r^\dag)=\sum_{r} \Tr(U^{\otimes n}M_r^\dag M_r \sigma)=  \Tr(U^{\otimes n} \sigma) \ ,
       \ee
proves the first part of the lemma.

To prove the second part, we consider the generalization of decomposition in Eq.(\ref{dec}) to $\mathrm{SU}(d)$ symmetry, namely the isotopic decomposition 
\be
(\mathbb{C}^d)^{\otimes n} \cong \bigoplus_{\lambda} \big(\mathbb{C}^{d_\lambda}\otimes \mathbb{C}^{m(n,\lambda)}\big)\ ,
\ee
where, according to Schur-Weyl duality, 
distinct $\lambda$ correspond to inequivalent irreps of $\mathbb{S}_n$, or equivalently, $\mathrm{SU}(d)$. Namely,   $\mathrm{SU}(d)$ acts irreducibly on $\mathbb{C}^{d_\lambda}$ and trivially on $\mathbb{C}^{m(n,\lambda)}$, and $\mathbb{S}_n$ acts irreducibly on $\mathbb{C}^{m(n,\lambda)}$ and trivially on $\mathbb{C}^{d_\lambda}$.

With respect to this decomposition, any PI state $\sigma$ takes the form
\be\nonumber
\sigma=\bigoplus_\lambda (\sigma_\lambda\otimes \frac{\mathbb{I}_{m(n,\lambda)}}{m(n,\lambda)})\ .
\ee
Similarly,
\be
U^{\otimes n}=\bigoplus_\lambda (u_\lambda\otimes \mathbb{I}_{m(n,\lambda)})\  ,
\ee
where $u_\lambda$ is the representation of $U$ on $\mathbb{C}^{d_\lambda}$.  On the other hand, any $\mathrm{SU}(d)$-invariant operator takes the form
\be
M_r=\bigoplus_\lambda(\mathbb{I}_{d_\lambda}\otimes \widetilde{M}_{r,\lambda})\ ,
\ee
where $\widetilde{M}_{r,\lambda}$ acts on $\mathbb{C}^{m(n,\lambda)}$.  Then,  the post-measurement state for outcome $r$ is 
\be
\frac{M_r \sigma M_r^\dag}{\Tr(M_r^\dag M_r \sigma)}=\frac{1}{\Tr(M_r^\dag M_r \sigma)} \bigoplus_\lambda \left(\sigma_\lambda \otimes \frac{\widetilde{M}_{r,\lambda} \widetilde{M}_{r,\lambda}^\dag }{m(n,\lambda)}\right)\ .
\ee
Assuming $\sigma$ is restricted to a single $\lambda$ sector, we find that in the above summation only one of the terms is non-zero, which implies
\be
\Tr\left(U^{\otimes n}\frac{M_r \sigma M_r^\dag}{\Tr(M_r^\dag M_r \sigma)}\right)=\frac{1}{\Tr(\widetilde{M}_{r,\lambda}^\dag\widetilde{M}_{r,\lambda})}  \Tr(u_\lambda \sigma_\lambda)  \Tr(\widetilde{M}_{r,\lambda} \widetilde{M}_{r,\lambda}^\dag) = \Tr(u_\lambda\sigma_\lambda)=\Tr(U^{\otimes n}\sigma)\ .
\ee
Then, it is straightforward to see that the original PI state $\sigma$ can be recovered by simply applying a random permutation, chosen uniformly at random, to the post-measurement state.

\end{proof}

\newpage

\section{Error Analysis (Proofs of Eq.(\ref{T-bound}) and Eq.(\ref{final-bound}))} \label{Appendix:D}

In this section, we finally prove the bound stated in Eq.~(\ref{final-bound}), namely
\begin{align}\label{rr}
D\big(\mathcal{E}_{\text{Schur}}(\sigma),\ \mathcal{E}^T_{\text{SWAP}}\left(\sigma \otimes \left|\tfrac{n}{2}\rangle\langle \tfrac{n}{2}\right|\right)\big) \le n \exp\left(-\tfrac{T}{2n}\right)\ ,
\end{align}
which also implies Eq.~(\ref{T-bound}). Recall from Eq.~(\ref{TD}) that for a PI state $\tau_j$ restricted to a sector with angular momentum $j$, we have
\begin{equation}
D\big(\mathcal{E}_{\text{Schur}}(\tau_j),\  \mathcal{E}^T_{\text{SWAP}}\left(\tau_j \otimes \left|\tfrac{n}{2}\right\rangle\left\langle \tfrac{n}{2}\right|\right)\big) = \Pr(j' \neq j; T)\ ,
\end{equation}
where $\Pr(j' \neq j; T)$ denotes the probability that the number of detected singlets, i.e., $n/2 - j'$, differs from the actual number of singlets in the system, namely $n/2 - j$. Equivalently, this is the probability that $j'>j$ after $T$ rounds of testing (Note that this could alternatively be expressed as $\Pr(j'\neq j|j, T)$, and this notation is used in Appendix ~\ref{App:Fid}). Next, observe that a general PI state $\sigma$ can be written as the convex combination of states restricted to angular momentum $j$ sectors
\be
\sigma=\sum_j p_j \tau_j.
\ee
Then, the triangle inequality implies 
\be
D\big(\mathcal{E}_{\text{Schur}}(\sigma)\ ,  \mathcal{E}^T_{\text{SWAP}}\left(\sigma \otimes \left|\tfrac{n}{2}\right\rangle\left\langle \tfrac{n}{2}\right|\right)\big)\le \sum_{j} p_j D\big(\mathcal{E}_{\text{Schur}}(\tau_j)\ ,  \mathcal{E}^T_{\text{SWAP}}\left(\tau_j \otimes \left|\tfrac{n}{2}\right\rangle\left\langle \tfrac{n}{2}\right|\right)\big)\ .
\ee
Indeed, in this case, because for different $j$, both $\mathcal{E}_{\text{Schur}}(\tau_j)$  and $ \mathcal{E}^T_{\text{SWAP}}\left(\tau_j \otimes \left|\tfrac{n}{2}\right\rangle\left\langle \tfrac{n}{2}\right|\right)$ 
live in the subspace with angular momentum $j$, and for different $j$, these subspaces are all orthogonal, this bound holds as identity, i.e., 
\bes
\begin{align}
D\big(\mathcal{E}_{\text{Schur}}(\sigma)\ ,  \mathcal{E}^T_{\text{SWAP}}\left(\sigma \otimes \left|\tfrac{n}{2}\right\rangle\left\langle \tfrac{n}{2}\right|\right)\big)&=  \sum_{j} p_j D\big(\mathcal{E}_{\text{Schur}}(\tau_j)\ ,  \mathcal{E}^T_{\text{SWAP}}\left(\tau_j \otimes \left|\tfrac{n}{2}\right\rangle\left\langle \tfrac{n}{2}\right|\right)\big)\\ &=\sum_{j} p_j\times  \Pr(j'\neq j; T)\label{monotonicityschur}\\ &\le \max_{j}\ \Pr(j'\neq j; T)\ .\label{tracedistancebound}
\end{align}
\ees
{It is worth noting that, as $T$ increases, the number of detected singlets can only increase. Moreover, since $\Pr(j'<j; T) = 0$, it follows that $\Pr(j'=j; T)$ is monotonically increasing, which in turn implies that $\Pr(j' \neq j; T)$ is monotonically decreasing with $T$. Consequently, the trace distance  
\[
D\big(\mathcal{E}_{\text{Schur}}(\sigma), \mathcal{E}^T_{\text{SWAP}}\left(\sigma \otimes \left|\tfrac{n}{2}\right\rangle\left\langle \tfrac{n}{2}\right|\right)\big)
\]
is monotonically decreasing with $T$.}\\

In the remainder of this appendix, we study $\Pr(j'\neq j; T)$ and show that it is bounded by $n \exp\left(-\tfrac{T}{2n}\right)$, which, in turn implies Eq.(\ref{rr}).  To do this, we first find the number of rounds of testing necessary to make sure $\Pr(j'\neq j; T)$ is bounded by $\epsilon$, as shown in Eq.~\eqref{Tbound}.

As we established in Eq.(\ref{proboferror}), 
\be
\text{Pr}(j'\neq j; T)=\Pr(X_1 + \cdots + X_{\frac{n}{2}-j}>T)\ ,\label{probtail}
\ee 
where $X_r: r=1,\cdots \frac{n}{2}-j$ are random variables corresponding to the number of SWAP tests needed to detect the $r$th singlet.  
As we saw in Eq.(\ref{mean1}), $X_r$ for all $r=1,\cdots \frac{n}{2}-j$ are geometric random variables with mean
$\langle X_r\rangle= \frac{1}{e_{j'}(r)}$. The probabilities $e_{j'}(r)$ are discussed in Appendix \ref{Appendix:probabilities}.
It follows that
\be\nonumber
T_{\ast}(j)=\sum_{j'>j}^{n/2} \frac{1}{e_{j'}(j)}\le 2(n-j)+(2j-2)  \ln(\frac{n}{2}-j)\ ,
\ee
where the bound is shown in Appendix \ref{Appendix C}. \\

Next, to bound the tail of the probability distribution in Eq.(\ref{probtail}), we use the following lemma, adopted from Theorem 2.1 of \cite{jansontail}. This lemma follows from a standard Chernoff bound argument.

\begin{lemma} (Theorem 2.1 of \cite{jansontail})
Let $Z_1,\cdots, Z_k$ be independent but not necessarily identically distributed geometric random variables with the success parameters $p_k$. Then, for $T \ge \mu$,
\begin{align}
    \operatorname{Pr}(Z_1+\cdots+Z_k \geq T) \leq \exp{\left(-p^*\mu\left(\frac{T}{\mu}-\ln{\left(\frac{T}{\mu}\right)-1}\right)\right)}, \label{jansonconc}
\end{align}
where $p^*=\min_k p_k$ and $\mu= \sum_k \frac{1}{p_k}$.
\end{lemma}
This lemma implies that  for $T\geq T_*(j)$,
\begin{align}
   \Pr(j'\neq j; T)&= \Pr(X_1 + \cdots + X_{\frac{n}{2}-j} > T)\\  &=\Pr(X_1 + \cdots + X_{\frac{n}{2}-j}  \geq T+1)\\ &\leq \exp(-p_j^*T_*(j)(\frac{T+1}{T_*(j)}-\ln(\frac{T+1}{T_*(j)})-1))\label{D2}
\end{align}
 where
\be \label{defp*T*}
p_j^*=\min_{j'}e_{j'}(j), \quad T_*(j) =\sum_{j'} \frac{1}{e_{j'}(j)}. 
\ee
Below in Sec.\ref{subsecsteps}, we prove that for 
\be\label{B12}
T\ge T_*(j)+\frac{2}{p_j^*}\ln\frac{1}{\epsilon}-1. 
\ee
the expression in Eq.(\ref{D2}) is bounded by $\epsilon$, which in turn implies $\Pr(j'\neq j; T)\le \epsilon$. In Sec.\ref{subsecsteps}, we also prove that
 \be \label{B13}
\max_j \left(T_*(j)+\frac{2}{p_j^*}\ln\frac{1}{\epsilon}-1\right) \le 2n\ln\frac{n}{\epsilon}-1. 
 \ee
It follows that by choosing 
 \be \label{Tbound}
T= \left\lceil 2n\ln\frac{n}{\epsilon}\right\rceil -1 \ge \max_j \left(T_*(j)+\frac{2}{p_j^*}\ln\frac{1}{\epsilon}-1\right)\ ,
\ee
we obtain 
\begin{align}
D\big(\mathcal{E}_{\text{Schur}}(\sigma)\ ,  \mathcal{E}^T_{\text{SWAP}}\left(\sigma \otimes \left|\tfrac{n}{2}\right\rangle\left\langle \tfrac{n}{2}\right|\right)\big)\le  \max_{j}\ \Pr(j'\neq j; T)\le \epsilon \ .
\end{align}
In summary, this means 
\begin{align}
D\big(\mathcal{E}_{\text{Schur}}(\sigma)\ ,  \mathcal{E}^T_{\text{SWAP}}\left(\sigma \otimes \left|\tfrac{n}{2}\right\rangle\left\langle \tfrac{n}{2}\right|\right)\big)\le  n \exp\left(-\tfrac{T}{2n}\right) \ ,
\end{align}
provided that
\[
T \geq n \ln\!\left( \frac{n}{2} \right) 
+ \frac{n}{\ln\!\left( \frac{n}{2} \right) - \ln\!\left( \ln\!\left( \frac{n}{2} \right) \right)} \ge \max_j T_*(j)\ , 
\] 
where the right-hand side of the above expression is an upper-bound for  $\max_j T_*(j)$, as shown in Eq. \eqref{Eq:maxTj}.
This proves Eq.(\ref{T-bound}) and Eq.(\ref{final-bound}). Therefore, to complete the proof, we only need to show Eq.~\eqref{B12} and Eq.~\eqref{B13}. We prove these in the following subsection, in Eq. \eqref{T_nj bound} and Eq. \eqref{T_nbound} respectively.

\subsection{Upper bound on the number of steps required to achieve error $\epsilon$ (Proof of Eq.(\ref{B12}) and Eq.(\ref{B13})) } \label{subsecsteps}

We find a lower bound on $T$ that guarantees the right-hand side of Eq.~\eqref{D2} is bounded by $\epsilon$, i.e.,
\be
\exp(-pT_*(\frac{T+1}{T_*}-\ln(\frac{T+1}{T_*})-1))\le \epsilon\ ,
\ee
which means
\begin{align}\label{d4}
    \frac{T+1}{T_*}-\ln{\left(\frac{T+1}{T_*}\right)}\leq \frac{1}{T_*}\left( T_*+\frac{1}{p}\ln\frac{1}{\epsilon}\right).
\end{align}
Note that we are using the shorthand $T_*$ for $T_*(j)$ and $p$ for $p^*_j$ in the rest of this subsection. Next, we use the following property, which we prove in the following subsection:
\be \label{x-lnx ineq}
 \forall a,x>1:\  
 x\leq a +\ln{a}\ \ \Longrightarrow\ \ x-\ln{x} \leq a.
\ee
Substituting
\begin{align*}
    x=\frac{T+1}{T_*},\quad a=\frac{1}{T_*}\left( T_*+\frac{1}{p}\ln\frac{1}{\epsilon}\right)
\end{align*}
 into $x\leq a +\ln{a}$ and multiplying both sides by $T_*$, we obtain the following chain of inequalities, where we have assumed $T\geq T_*$, 
\begin{align*}
    T+1 &\leq T_*\left(\frac{1}{T_*}\left( T_*+\frac{1}{p}\ln\frac{1}{\epsilon}\right)+\ln\left(\frac{1}{T_*}\left( T_*+\frac{1}{p}\ln\frac{1}{\epsilon}\right)\right)\right) \\  
    &\leq T_*+\frac{1}{p}\ln\frac{1}{\epsilon}+T_*\ln\left( 1+\frac{1}{pT_*}\ln\frac{1}{\epsilon}\right) \\
    &= T_*+\frac{1}{p}\ln\frac{1}{\epsilon}+T_*\ln\left( 1+\frac{1}{pT_*}\ln\frac{1}{\epsilon}\right) \\
    &\leq T_*+\frac{1}{p}\ln\frac{1}{\epsilon}+\frac{1}{p}\ln\frac{1}{\epsilon}\ ,
\end{align*}
where we used the fact that $\ln(1+x) \leq x$ in the last step. Then, 
Then, Eq.(\ref{x-lnx ineq}) implies that for 
\begin{align}
    T= T_*(j)+\frac{2}{p_j^*}\ln\frac{1}{\epsilon}-1. \label{T_nj bound}
\end{align} 
condition in Eq. \eqref{d4} is satisfied. 
This proves the bound Eq.(\ref{B12}) implies  $\Pr(j'\neq j; T)\le \epsilon$.

Next, we prove Eq.(\ref{B13}).
We derive the following upper bound for the right-hand side of Eq.~\eqref{T_nj bound}:
\begin{align}
    &\max_{0\leq j \leq n/2-1} T_*(j) + \frac{2}{p_j^*}\ln\frac{1}{\epsilon}-1\\
    &\leq \max_{0\leq j \leq n/2} T_*(j) + \max_{0\leq j \leq \frac{n}{2}-1}\frac{2}{p_j^*}\ln\frac{1}{\epsilon}-1\\
    &\leq \left(n\ln{\frac{n}{2}}+\frac{n}{\ln\!\left(\frac{n}{2}\right) - \ln\!\!\left(\ln\!\frac{n}{2}\right)}\right) + 2(n-1)\ln\frac{1}{\epsilon}-1\ . \label{jd}
\end{align}
For the last step, we used the following inequality, derived in Appendices \ref{Appendix C}:
\begin{align}
     \max_{0\leq j \leq n/2}T_*(j) \leq \left(n\ln{\frac{n}{2}}+\frac{n}{\ln\!\left(\frac{n}{2}\right) - \ln\!\!\left(\ln\!\frac{n}{2}\right)}\right), \label{B21}
\end{align} 
and, 
\begin{align}
    \min_{0\leq j \leq \frac{n}{2}-1} &p_*(j)=\frac{1}{n-1}\ ,
\end{align}
which is shown below
\begin{align}
    \min_{0\leq j \leq \frac{n}{2}-1} \, p_*(j) 
    &= \min_{0\leq j \leq \frac{n}{2}-1} \, \min_{j+1\leq j'\leq \frac{n}{2}} \, e_{j'}(j) \\
    &= \min_{1\leq j' \leq \frac{n}{2}} \, \min_{0\leq j\leq j'-1} \, e_{j'}(j) \\
    &= \min_{1\leq j' \leq \frac{n}{2}} \, \min_{0\leq j\leq j'-1}
    \frac{j'(j'+1)-j(j+1)}{2j'(2j'-1)} \\
    &= \min_{1\leq j' \leq \frac{n}{2}}
    \frac{j'(j'+1)-(j'-1)j'}{2j'(2j'-1)} \\
    &= \min_{1\leq j' \leq \frac{n}{2}}
    \frac{2j'}{2j'(2j'-1)} 
    = \min_{1\leq j' \leq \frac{n}{2}}
    \frac{1}{2j'-1} \\
    &= \frac{1}{n - 1}.
    \label{B22}
\end{align}

By slightly relaxing the bound in Eq.(\ref{jd}), we obtain a more convenient expression, namely
\begin{align}
   &(n\ln{n}-n\ln{2}+\frac{n}{\ln\!\left(\frac{n}{2}\right) - \ln\ln\frac{n}{2}}) + 2(n-1)\ln\frac{1}{\epsilon}-1\\
&\le  n\ln{n} + 2(n-1)\ln\frac{1}{\epsilon}-1 \\ 
&\le 2n\ln{n} + 2n\ln\frac{1}{\epsilon}-1\\
&\le 2n\ln\frac{n}{\epsilon}-1. \label{T_nbound}
\end{align}
This completes the proof of Eq. \eqref{B13}.

\subsection{Proof of Eq. \eqref{x-lnx ineq}}
In this subsection, we provide a proof for Eq. \eqref{x-lnx ineq}, i.e., we show that for all $a>1$ and $x>1$, if $a +\ln{a} \geq x$, then $\quad x-\ln{x} \leq a$.  

Let 
\begin{align*}
    y=a +\ln{a}\ , 
\end{align*}
This implies that $y\geq x >1$. Note then, $x-\ln{x} \leq y-\ln{y}$. This is because for $z>1$,
\be
\frac{d}{dz} (z-\ln{z})=1-\frac{1}{z}>0,
\ee
i.e., $z-\ln{z}$ increases monotonically. Then,  
\begin{align}
    x-\ln{x}&\leq y-\ln{y} \\
    &= a+\ln{a} - \ln{y}\\ 
    &= a- \ln{(\frac{y}{a})}\\ 
    &\leq  a.   
\end{align}
The last inequality is true because $y=a+\ln{a}$ implies that $y \geq a$ for all $a> 1$.

\newpage

\color{black}

\section{Probability of finding a singlet  (Proof of Eq.(\ref{cond}))} \label{Appendix:probabilities}

Suppose we start with state $\tau_j$ defined in {Eq.(\ref{input})}. In this section, we calculate the probability $\text{P}(k+1 ; T+1 \mid k ; T)$ that the $(T+1)$'th SWAP test detects the $(k+1)$-th singlet, given that $k$ singlets have been detected after $T$ SWAP tests. Let $\omega_{A_k}$ be the state of qubits in $A_k$ as defined in Eq.(\ref{def}).  That is,
\begin{align}
\text{P}(k+1 ; T+1 \mid k ; T)&=\frac{\Tr(\mathcal{S}_{A_{k}}(\omega_{A_{k}}))}{\Tr(\omega_{A_{k}})}\\  &=
 \frac{1}{\binom{n-2k}{2}} \times \sum_{r< s} \Tr\left(\xi_{r,s} \frac{\omega_{A_{k}}}{\Tr(\omega_{A_{k}})}   \right)\\  &=
 \Tr\left(E_{j'} \frac{\omega_{A_{k}}}{\Tr(\omega_{A_{k}})}   \right)\ ,
\end{align}
where $j'=n/2-k$, and 
\begin{align}
E_{j'}&=\frac{1}{\binom{2j'}{2}}\sum_{r < s} \xi_{r,s}= \frac{1}{\binom{2j'}{2}}\sum_{r < s} \frac{\mathbb{I}-S_{r,s}
}{2}=\frac{\mathbb{I}}{2}-\frac{1}{2\binom{2j'}{2}}\sum_{r<s} S_{r,s}.
\end{align}
For a system with $2j'$ qubits, the average of all SWAP operators on all pairs of qubits is  
\bes
\begin{align}
\frac{1}{\binom{2j'}{2}}\sum_{r<s} S_{r,s}&=\frac{1}{\binom{2j'}{2}}\sum_{r<s}\frac{\mathbb{I}+\vec{\sigma}_r\cdot \vec{\sigma}_s}{2}\\ &=\frac{\mathbb{I}}{2}+ \frac{1}{2\binom{2j'}{2}}\times \frac{4 J^2-6j' \mathbb{I}}{2}\\ &=\frac{1}{\binom{2j'}{2}}  J^2+\Bigg[\frac{1}{2}-\frac{3j'}{2\binom{2j'}{2}}\Bigg] \mathbb{I}\ ,
\end{align}
\ees
\color{black}
 where $J^2$ is the operator $J^2=J_x^2+J_y^2+J_z^2$ defined on $2j'$ qubits and $J_w=\sum_{r=1}^{2j'} \sigma^{(w)}_r/2: w=x,y,z$. This implies that
\bes
\begin{align}
E_{j'}&=\frac{\mathbb{I}}{2}-\frac{1}{2\binom{2j'}{2}}\sum_{r < s} S_{r,s}=\Bigg[\frac{1}{4}+\frac{3j'}{4 \binom{2j'}{2}}\Bigg] \mathbb{I}-\frac{1}{2\binom{2j'}{2}}  J^2.
\end{align}
\ees
 Therefore,
\begin{align}
E_{j'}&=\sum_{l} e_{j'}(l)\Pi_l\ ,
\end{align}
where $\Pi_l$ is the projector to the eigensubspace of $J^2$ with eigenvalue $l(l+1)$ and 
\be
e_{j'}(l)=\Bigg[\frac{1}{4}+\frac{3j'}{4 \binom{2j'}{2}}\Bigg] -\frac{1}{2\binom{2j'}{2}} l(l+1)= \frac{j'(j'+1)-l(l+1)}{2j'(2j'-1)}\  .
\ee
Since state ${\omega_{A_{k}}}/{\Tr(\omega_{A_{k}})} $ is restricted to the sector with angular momentum $j$, 
we have $\Pi_l {\omega_{A_{k}}}/{\Tr(\omega_{A_{k}})} =\delta_{l,j} {\omega_{A_{k}}}/{\Tr(\omega_{A_{k}})}$, which in turn implies
\begin{align}
\text{P}(k+1 ; T+1 \mid k ; T)&=\sum_le_{j'}(l) \Tr(\Pi_l \frac{\omega_{A_{k}}}{\Tr(\omega_{A_{k}})})\label{B9}\\
&=e_{j'}(j) \Tr(\Pi_j \frac{\omega_{A_{k}}}{\Tr(\omega_{A_{k}})})=e_{j'}(j)=\frac{j'(j'+1)-j(j+1)}{2j'(2j'-1)}\ ,
\end{align}
\color{black}
where $j'=n/2-k$. This proves Eq.(\ref{cond}). Note that this conditional probability is independent of $T$, which implies that the random process of detecting the $k$-th singlet follows a geometric distribution.

\subsection{The minimum probability for each $j$}

In this subsection, we calculate the minimum over all $j'$ in the interval $j+1\le j'\le n/2$ of $\text{P}(k+1 ; T+1 \mid k ; T)=e_{j'}(j)$ for a fixed $j$. That quantity, which we have previously defined as $p_*(j)$ in Eq. \eqref{defp*T*}, is
\be
p_\ast(j)=\min_{j+1\le j'\le n/2} e_{j'}(j)= \min_{j+1\le j'\le n/2} \frac{j'(j'+1)-j(j+1)}{2j'(2j'-1)}\ .
\ee
In the context of the Markov chain described in Fig.~\ref{fig:enter-label}, we find the state $j'$ in the Markov chain that is ``hardest'' to escape besides the fixed point $j$. Note that we need to consider $j$ values that are non-negative integers (for even $n$) and positive half integers (for odd $n$). In particular, we show that 
\be
p_*(j)=\min_{j+1\le j'\le n/2} \, e_{j'}(j) =
\begin{cases}
\dfrac{n+2}{4(n-1)} & \text{if } j = 0. \\[1.5ex]
\dfrac{1}{2j+1} & \text{if } j > 1.\\[1.5ex]
\min\left( \dfrac{1}{3}, \dfrac{1}{4} + \dfrac{8}{n} - \dfrac{5}{4} \cdot \dfrac{1}{n-1} \right) & \text{if } j = 1. \\
\frac14+\frac{3}{4n} & \text{if } j = \frac12.
\end{cases}
\ee

For $j>1$, the minimum of $e_{j'}(j)$ is attained at $j'=j+1$. This is the penultimate step in the  Markov chain in Fig.~\ref{fig:enter-label}. For $j=0$, on the other hand, the minimum is attained for $j'=n/2$. For $j=1$ the minimum is attained either at $j'=2$ or $j'=n/2$. \\

First, note that
\begin{align*}
e_{j'}(j)=\frac{(j' - j)\,(j' + j + 1)}{2 j' (2 j' - 1)}
  =
  \frac{1}{4}\!\left(
      1
      + \frac{j\,(2j + 2)}{j'}
      - \frac{2j^2 +2j - 1.5}{\,j' - 0.5}
  \right).
\end{align*}
We break this into four cases: 
\begin{itemize}
    \item $j=0$ (when n is even):  In this case we show that $\min_{j'}e_{j'}(0)=\frac{n+2}{4(n-1)}$ and the minimum is attained at $j'=n/2.$
    \color{black}    We find that $e_{j'}(0)=\frac14(1+\frac{1.5}{j'-0.5})$, so 
    \begin{align}
       p_*(0)=\min_{1\leq j' \leq n/2}\frac14\left(1+\frac{1.5}{j'-0.5}\right)=\frac14\left(1+\frac{3}{n-1}\right)=\frac{n+2}{4(n-1)},
    \end{align}
     where the minimum is attained at $j'=n/2.$ Note that $p_*(0)=\frac{n+2}{4(n-1)} \geq \frac14.$
    \item $j>1:$  In this case we show that $$\min_{j'}e_{j'}(j)=\frac{1}{2j+1}$$ and the minimum is attained at $j'=j+1.$
    \color{black} We first note that in the range $[j+1, \infty)$, the function 
    \begin{align*}
        f(x)=\frac{1}{4}\!\left(
      1
      + \frac{j\,(2j + 2)}{x}
      - \frac{2j^2 +2j - 1.5}{\,x - 0.5}
  \right)
  \end{align*}
  has only one local maximum. To see this, let $x^*$ be the point where the first derivative is 0. That is,
  \begin{align*}
      &f'(x^*)=-\frac{j\,(2j + 2)}{4x^{*2}}
      + \frac{2j^2 +2j - 1.5}{\,4(x^* - 0.5)^2}=0 \ .
  \end{align*}
  One of the solutions to the above equation is
  \begin{align*}     x^*=\frac13\left(2j^2+2j\;+\;\sqrt{\,(2j^2+2j)(2j^2 + 2j - 1.5)\,}\right)\ge j+1 \ .
  \end{align*}
  Hence, $ (\frac{1}{x^*}-\frac{1}{x^*-5})\le 0.$
  Therefore, its second derivative at $x^*$ is negative:
\begin{align*}
      &f''(x^*)=\frac{j\,(2j + 2)}{2x^{*3}}
      - \frac{2j^2 +2j - 1.5}{\,2(x^* - 0.5)^3}=\frac{j\,(2j + 2)}{2x^{*2}}(\frac{1}{x^*}-\frac{1}{x^*-0.5})\le 0.
  \end{align*}
  where we used the fact that $f'(x^*)=0$. We disregard the other solution because it is $$\frac13(2j^2+2j\;-\;\sqrt{\,(2j^2+2j)(2j^2 + 2j - 1.5)\,})\le j+1$$ 
  and is therefore not in the range. Thus, as $x$ increases towards $x^*$, the function approaches its local maximum. Afterwards, it begins to decrease monotonically to $\frac14$ since 
  \begin{align*}
      \lim_{x \rightarrow \infty}\frac{1}{4}\!\left(1
      + \frac{j\,(2j + 2)}{x}
      - \frac{2j^2 +2j - 1.5}{\,x - 0.5}
  \right) =\frac14.
  \end{align*}
  Since $f(j+1)=\frac{1}{2j+1}\leq \frac14$ for all $j>1$, it follows that $\min_{x \geq j+1}f(x)=\frac{1}{2j+1}$. As a consequence,
  \begin{align}
       &\min_{j+1\le j'\le n/2} \frac{(j'-j)(j'+j+1)}{2j'(2j'-1)}=\frac{1}{2j+1}.
  \end{align}
We have shown that 
\be
    \min_{j' \in \{2,n/2\} }e_{j'}(j)=\min_{j+1 \le x \le n/2}\frac{1}{4}\!\left(1
      + \frac{j\,(2j + 2)}{x}
      - \frac{2j^2 +2j - 1.5}{\,x - 0.5}
  \right)=\frac1{2j+1}
\ee
which is achieved at $x^*=j+1$. It follows that $p_*(j)=\frac{1}{2j+1}$ for all $j > 1$.
Note that for $j>1$, $p_*(j)=\frac{1}{2j+1}\geq \frac1{n+1}$.
  
 \item $j=1$ (when n is even) :  In this case, we show that $$\min_{j'}e_{j'}(1)=\min{\{e_{2}(1),e_{n/2}(1)\}}=\min{\{\frac13,(\frac14+\frac8n-\frac54.\frac1{n-1})\}}\geq \frac14.$$ 
    Note that
 \begin{align*}
     e_{j'}(1)=\frac{1}{4}\!\left(
      1
      + \frac{4}{j'}
      - \frac{2.5}{\,j' - 0.5}
  \right).
 \end{align*}
 
 Once more, consider the continuous version of this function 
 \be
 f(x)=\frac{1}{4}\!\left(
      1
      + \frac{4}{x}
      - \frac{2.5}{\,x - 0.5}\right).
\ee
      This function has only one local maximum at $x^* = \frac{4}{3} + \sqrt{\frac{10}{9}}$ for all $n > 4$. Then the minimum of $f(x)$ will be at one of the two endpoints. It follows that $e_{j'}(1)$ will have its minimum either at $j'=2$ or $j'=n/2$, so
 \begin{align}
       &\min_{j' \in \{2,n/2\} }e_{j'}(1)=\min_{j' \in \{2,n/2\} }\frac{(j'-1)(j'+2)}{2j'(2j'-1)}=\min{\{\frac13,(\frac14+\frac8n-\frac54.\frac1{n-1})\}} \geq \frac14.
  \end{align}
  
    \item $j=\frac12$ (when $n$ is odd): In this case
    \begin{align*}
        e_{j'}(\frac12)=\frac14 + \frac{3}{8j'}.
    \end{align*}
    Then 
    \begin{align}
        &\min_{j' \in \{3/2,n/2\} }e_{j'}(\frac12)=\frac14+\frac{3}{4n}>\frac14.
    \end{align}
 
\end{itemize}

 \color{black}
 \newpage

\section{Upper and Lower bounds on the mean number of SWAP tests (Proof of Eq.(\ref{eqn:15})} \label{Appendix C}

\color{black}

In this section, we establish bounds for the mean number of SWAP tests  $T_\ast(j)$ presented in Eq.(\ref{eqn:15}). In addition, we prove that 
\begin{align} \label{Eq:maxTj}
    \max_{0 \le j \le n/2} T_*(j) 
\leq n \ln\!\left( \frac{n}{2} \right) 
+ \frac{n}{\ln\!\left( \frac{n}{2} \right) - \ln\!\left( \ln\!\left( \frac{n}{2} \right) \right)}\ ,
\end{align}

in subsection~\ref{maxTj} . Recall that
\begin{align*}
T_{\ast}(j) 
= \sum_{j' > j}^{n/2} \frac{1}{e_{j'}(j)}
= \sum_{j' = j+1}^{n/2} \frac{2j'(2j'-1)}{(j'-j)(j'+j+1)}\ . 
\end{align*}
We used Eq. \eqref{prob} for the last equality. Expanding $\frac{1}{e_{j'}(j)}$, we find that
\be \label{c1}
\frac{1}{e_{j'}(j)}=\frac{2j'(2j'-1)}{(j'-j)(j'+j+1)}
= 4+\frac{2j-2}{j'-j}-\frac{2j+4}{j'+j+1}+ \frac{2}{(j'-j)(j'+j+1)}\ .
\ee
This implies
\be
 4+\frac{2j-2}{j'-j}-\frac{2j+4}{j'+j+1} \le \frac{1}{e_{j'}(j)}\le  4+\frac{2j-2}{j'-j}\ .
\ee
The lower bound follows from ignoring the last term in \ref{c1}, while the upper bound follows from the fact that 
\begin{align*}
    \frac{2}{(j'-j)(j'+j+1)}-\frac{2j+4}{j'+j+1}  \leq 0.
\end{align*}
This, in turn, implies the upper bound
\begin{align}\label{T(j) upper bound}
T_\ast(j)= \sum_{j' > j}^{n/2} \frac{1}{e_{j'}(j)} &\leq \sum_{j' = j+1}^{n/2}4+ \frac{2j -2}{j'-j} 
= 2(n - 2j) 
+ (2j-2) \left( 1 + \frac{1}{2} + \cdots + \frac{1}{n/2 - j} \right)
& \\[6pt]
&
\le 2n - 2j 
+ (2j-2)\ln(n/2 - j)\ , 
&
\end{align}
where the last line follows from
\be \label{logineq}
\ln\frac{b}{a-1}\geq \frac1a+\frac1{a+1}+ \cdots +\frac{1}{b} \geq \ln{\frac{b+1}{a}}.
\ee
We provide justification for this in the footnote below\footnote{This follows from $    \int_{a-1}^b \frac{1}{x} \, dx \geq \sum_{k=a}^{b} \frac{1}{k} \geq \int_{a}^{b+1} \frac{1}{x} \, dx. $
}. This completes the proof for Eq. \eqref{eqn:15}.
We also provide a lower bound for $T_*(j)$, 
\begin{align}
   T_\ast(j)&= \sum_{j' > j}^{n/2} \frac{1}{e_{j'}(j)} \ge \sum_{j' = j+1}^{n/2}4+ \frac{2j-2}{j'-j}-\frac{2j+4}{j'+j+1}\\
    &\geq 2n-4j+ (2j-2)\ln(n/2 - j+1)- \sum_{j' = j+1}^{n/2} \frac{2j+4}{j'+j+1} \ .\label{c7}
\end{align}
Using the inequality in Eq. \eqref{logineq} to bound the summation in the right-hand side of Eq.(\ref{c7}), we find 
\begin{align*}
    &\sum_{j' = j+1}^{n/2}\frac{2j+4}{j'+j+1}
    \le (2j+4)\ln\left(\frac{\frac{n}{2}+j+1}{2j+1}\right),
\end{align*}
which implies 
\be \label{T(j) stronger lower bound}
    T_*(j) \ge 2n-4j + (2j-2)\ln(n/2 - j)-(2j+4)\ln\left(\frac{\frac{n}{2}+j+1}{2j+1}\right)\ .
\ee
The behavior of $T_*(j)$ and these bounds are illustrated in the bottom plot in Fig. \ref{fig:probj'neqj}. Note that the gap between the upper and lower bounds in Eq.(\ref{T(j) upper bound}) and Eq.(\ref{T(j) stronger lower bound}) is 
\begin{align} 
      2j+(2j+4)\ln(\frac{\frac{n}{2}+j+1}{2j+1})\ ,
\end{align}
We find a bound for this gap that's independent of $j$:
\begin{align}
    2j+(2j+4)\ln(\frac{\frac{n}{2}+j+1}{2j+1}) &\le 2j + (\frac{2j+4}{2j+1}+\frac{2j+4}{2j+2}+\frac{2j+4}{2j+3}+\frac{2j+4}{2j+4}+ \cdots +\frac{2j+4}{n/2-j})\\
    &\le2j+(4+2+\frac{4}{3}+1+\cdots+1)\le 2j+\frac{n}{2}-j+5\\
    &\le n+5\ ,\label{T gap}
\end{align}
where the first inequality follows from  Eq.(\ref{logineq}).\\

\color{black}
We use the above results to study the behavior of ${\min_{j'} e_{j'}(j)}, T_*(j)$ and $\eta_j=p_j^*\times T_*(j)$ in different regimes of $j$ for large values of $n$. The following table summarizes these results:

\begin{table}[h!]
\centering
{
\renewcommand{\arraystretch}{1.3} 
\setlength{\tabcolsep}{20pt}      

\begin{tabular}{|l|c|c|c|}
\hline
 & \textbf{$\displaystyle {\min_{j'} e_{j'}(j)}$} 
 & \textbf{$\displaystyle T_*(j)$} 
 & \textbf{$\displaystyle \eta_j$} \\ \hline

$\displaystyle j = c$ 
& $\displaystyle \frac{1}{2c+1}$ 
& $\displaystyle \alpha_1 n$ 
& $\displaystyle \frac{\alpha_1 n}{2c+1}$ \\ \hline

$\displaystyle \frac{2j}{n} = c$ 
& $\displaystyle \frac{1}{cn}$ 
& $\displaystyle cn\ln n $ 
& $\displaystyle \ln n $ \\ \hline

$\displaystyle j = \frac{n}{2} - c$ 
& $\displaystyle \frac{1}{\,n - 2c + 1\,}$ 
& $\displaystyle \alpha_2 n$ 
& $\displaystyle 2\alpha_2$ \\ \hline

\end{tabular}
}
\caption{Leading order terms in $n$ for ${\min_{j'} e_{j'}(j)}, T_*(j)$ and $\eta_j$ in the three regimes of $j$, where $c=o(\ln(n))$. The unknown constants $\alpha_1, \alpha_2$ come from the gap in the bounds of $T_*(j)$, which can be bounded by $n+5$, as shown  in Eq.~\eqref{T gap}.}
\label{tab:example}
\end{table}

\subsection{Proof of Eq.~\eqref{Eq:maxTj}}\label{maxTj}
In this subsection, we prove the upper bound on $\max_j T_*(j)$   stated in Eq. \eqref{Eq:maxTj}. We restate it here for convenience.
\begin{align*}
\max_{0 \le j \le n/2} T_*(j) 
\leq n \ln\!\left( \frac{n}{2} \right) 
+ \frac{n}{\ln\!\left( \frac{n}{2} \right) - \ln\!\left( \ln\!\left( \frac{n}{2} \right) \right)}\ .
\end{align*}
We start with the following upper bound proved in the previous section
\begin{align}
\max_{0 \le j \le n/2} T_*(j) 
&\le \max_j \left( 2n - 2j + (2j - 2) \ln\left( \frac{n}{2} - j \right) \right) \notag \\
&\le \max_{0 \le x \le \frac{n}{2}} \left( 2n - 2x + 2x \ln\left( \frac{n}{2} - x \right) \right). \label{c8}
\end{align}
We maximize the right-hand side by differentiating and setting the result equal to zero, which yields
\be \label{derivativeinC}
    \left(\frac{n}{2}-x^*\right)\ln\!\left( \frac{n}{2}-x^* \right)=\frac{n}{2}.
\ee
Taking the logarithm on both sides of Eq. \eqref{derivativeinC} and rearranging, we find that
\be \label{c14}
\ln\!\left( \frac{n}{2}-x^* \right) =\ln\!\left( \frac{n}{2} \right)- \ln\!\left( \ln\!\left( \frac{n}{2}-x^* \right) \right) \ge \ln\!\left( \frac{n}{2} \right)- \ln\!\left( \ln\!\left( \frac{n}{2} \right) \right).
\ee
Then, using Eq.\eqref{c14} to bound the left-hand side of the expression, we obtain
\be \label{c19}
   \frac{\frac{n}{2}}{\ln(\frac{n}{2})-\ln\!\left( \ln\!\left( \frac{n}{2} \right) \right)} \ge \frac{n}{2}-x^*=\frac{\frac{n}{2}}{\ln(\frac{n}{2}-x^*)} \ge  \frac{\frac{n}{2}}{\ln(\frac{n}{2})}.
\ee
Equivalently,
\be \label{c20}
   \frac{n}{2}\Bigg(1-\frac{1}{\ln(\frac{n}{2})-\ln\!\left( \ln\!\left( \frac{n}{2} \right) \right)}\Bigg) \le x^* \le \frac{n}{2}\Bigg(1- \frac{1}{\ln(\frac{n}{2})}\Bigg),
\ee
so the maximum is attained at $j^* \approx x^*$. Substituting Eq. \eqref{c19} and Eq. \eqref{c20} into Eq. \eqref{c8} and omitting the negative terms, we get the desired upper bound stated in Eq. \eqref{Eq:maxTj}.\\
\begin{align*}
\max_j T_*(j) \leq n \ln\!\left( \frac{n}{2} \right) 
+ \frac{n}{\ln\!\left( \frac{n}{2} \right) - \ln\!\left( \ln\!\left( \frac{n}{2} \right) \right)}.
\end{align*}

\newpage
\section{Average Fidelity of purification (Proof of Eq.(\ref{fid}))}\label{App:Fid}
A purification protocol based on random SWAP tests was proposed in the paper. Specifically, the protocol is applied to $n$ copies of a noisy qubit state $\rho$, and one of the remaining qubits not discarded by the SWAP test is returned. Any singlet that is detected in the process is moved away (see Fig.~\ref{Fig}).
In this section, we analyze the expected fidelity of the qubit returned after $T$ rounds of random SWAP testing on $n$ copies of the input state $\rho$. 
Due to the $\mathrm{SU(2)}$ symmetry of the protocol, we may assume $\rho=(1-p)|0\rangle\langle 0|+p\mathbb{I}/2$, where $0\leq p \leq 1$ without loss of generality (a similar argument is made and explained in detail in \cite{Cirac}). 

We show that the average fidelity attained after $T$ rounds of testing on $n$ copies of $\rho$ is 
\begin{align}  f(n,T)=\mathbb{E}_{j,j'}\Bigg[\frac{1}{2}+\frac{j}{j'}(f_j-\frac12)\Bigg]
    =\frac12 + \sum_{j,j'}\frac{j}{j'}(f_j-\frac12)\times p_j\times \text{P}(j'|j,T) \ ,\label{avgfid}
\end{align}
where, $ p_j $ and $f_j$ can be defined in terms of Schur transform of $\rho^{\otimes n}$, namely 
\be
\mathcal{E}_{\text{schur}}(\rho^{\otimes n})=\sum_{j}p_j\ \rho_j \otimes \xi^{\otimes n-2j}  \otimes |j\rangle\langle j|_{\text{reg}}\ .
\ee
Then, $ p_j $ is the probability that the system is found in the subspace with angular momentum $j$, and  $f_j$ is the fidelity of each qubits in state $\rho_j$ with the desired state $|0\rangle$. That is,
\be\label{mb}
f_j=\langle 0| \Tr_{\bar{1}}(\rho_j)|0\rangle=\frac{1}{2}+ \frac{\Tr(\rho_j J^{(j)}_z)}{2j}  \ ,
\ee
where $\Tr_{\bar{1}}(\rho_j)$ is the reduced state of the first qubit, 
\be
J^{(j)}_z=\frac{1}{2}\sum_{r=1}^{2j}Z_r\ ,
\ee
and  we have used the fact that $|0\rangle\langle 0|=(\mathbb{I}+Z)/2$. Both $f_j$ and $ p_j $  are calculated in \cite{Cirac}, and shown to be 
\begin{align}
    p_j &= (1-p)^{-1}m(n,j) \left[ \frac{2p-p^2}{4} \right]^{\frac{n}{2} - j}
    \cdot
    \Big[\left(1- \frac{ p}{2} \right)^{2j+1} - \left( \frac{ p}{2} \right)^{2j+1}\Big]
    \ ,\\
   f_j &= \frac{1}{2j} \left[
\frac{(2j + 1) \left(1 - \frac{p}{2} \right)^{2j+1}}{
\left(1 - \frac{p}{2} \right)^{2j+1} - \left( \frac{ p}{2} \right)^{2j+1}
}
- \frac{1-\frac{p}{2}}{1-p}
\right].
\end{align}
  where $m(n,j) = \binom{n}{\frac{n}{2} - j} - \binom{n}{\frac{n}{2} - j - 1}$ is the multiplicity of angular momentum $j$.\\


Note that the average purification fidelity via the Schur transform can also be obtained  for the formula in Eq.(\ref{avgfid}) by choosing 
$\text{P}(j'|j,T)=\delta_{j,j'}$, which achieves
\begin{align}  f_n^\text{opt}=
   \sum_{j}  p_j f_j \ .
\end{align}

Using the fact that,  $\text{P}(j'|j,T)$ is zero for $j'<j$ we find that
\begin{align}
f_n^\text{opt}- f(n,T)&= \sum_{j,j'} \left[\delta_{j,j'}-\frac{j}{j'}\text{P}(j'|j,T)\right]\times (f_j-\frac12)\times p_j \label{Eq: fiddiff}\ . 
\end{align}

Let 
\be
\epsilon= \max_{j} \text{P}(j'\neq j; T)=\max_{j} \sum_{j'\neq j}\text{P}(j'|j,T)
\ee

This implies that for all $j$
\[
    \text{P}(j'=j|j,T) \ge 1-\epsilon.
\]
Combining this with $\text{P}(j'=l|j,T) \ge 0$ for all $l\neq j$, we have
\[
    \text{P}(j'|j,T) \ge \delta_{j,j'}\times (1-\epsilon).
\]

Using this in Eq. \eqref{Eq: fiddiff},
\begin{align}
    f_n^\text{opt}- f(n,T)\leq &\sum_{j,j'} \left[\delta_{j,j'}-\frac{j}{j'}\delta_{j,j'}\times(1-\epsilon)\right]\times (f_j-\frac12)\times p_j\\
    &\leq \sum_{j} \left[1-\frac{j}{j}(1-\epsilon)\right]\times (f_j-\frac12)\times p_j\\
    &=\epsilon \times \sum_j  (f_j-\frac12)\times p_j\\
    &=\epsilon\times (f^{\text{opt}}_n-\frac12).
\end{align}

To prove Eq.(\ref{avgfid}), consider the following class of purification protocols, which includes our SWAP-test-based protocol as well as Schur transforms as special cases: First, a number of singlets are detected in the system, e.g., via a SWAP test, resulting in a state of the form
\be
\mathcal{E}(\rho^{\otimes n})=\sum_{j}  \beta_j \otimes \xi^{\otimes n/2-j}\otimes |j\rangle\langle j|\ , 
\ee
where $\beta_j$ is the unnormalized state of the remaining qubits when $n/2-j$ singlets are detected.   Then, we return one of the remaining qubits in state $\beta_j$ chosen uniformly at random.  For any such protocol,  the fidelity with the desired output state $|0\rangle$ can be written as
\begin{align}
\text{Fid}(\mathcal{E})= \sum_{j}  \frac{1}{2j}\sum_{r=1}^{2j} \Tr\left(\beta_j |0\rangle\langle 0|_r\right) &=\sum_{j}   \frac{1}{2j}\sum_{r=1}^{2j} \Tr\left(\beta_j \left(\frac{\mathbb{I}+Z_r}{2}\right)\right)\\ 
&=\frac{1}{2}+\sum_{j} \frac{\Tr(\beta_j J^{(j)}_z)}{2j} \ ,
\end{align}
where $|0\rangle\langle0|_r=(\mathbb{I}+Z_r)/2$ is the projector $|0\rangle\langle 0|$ at qubit $r$ tensor product with the identity operators on the rest of the qubits.

Define  $\beta_{j'|j} $ as the output of the protocol $\mathcal{E}$ on input $\mathcal{P}(\rho_j\otimes \xi^{\otimes (n/2-j)})$.
\be
\mathcal{E}\circ\mathcal{P}(\rho_j\otimes \xi^{\otimes (n/2-j)})=\sum_{j'}  \beta_{j'|j}\otimes \xi^{\otimes (n/2-j')}\otimes |j'\rangle\langle j'| \ ,
\ee
which means
\be
\beta_{j'}=\sum_{j} p_j  \beta_{j'|j}  \ ,
\ee
which, in turn, implies 
\begin{align}\label{cm}
\text{Fid}(\mathcal{E})= \frac{1}{2}+\sum_{j, j'} p_j \frac{\Tr(\beta_{j'|j} J^{(j')}_z)}{2j'} \ ,
\end{align}
Assuming the channel $\mathcal{E}$ respects the strong $\mathrm{SU}(2)$ symmetry, 
\be
\mathcal{E}(U^{\otimes n}\cdot)=U^{\otimes n} \mathcal{E}(\cdot)\ \ ,\ \  \mathcal{E}(\cdot U^{\otimes n})= \mathcal{E}(\cdot)U^{\otimes n}\ ,
\ee
for all $U\in\mathrm{SU}(2)$, then applying lemma \ref{lem:SU} to PI state $\sigma=\mathcal{P}(\rho_j\otimes \xi^{\otimes (n/2-j)})$, we find
\be
\Tr\left(J_z^{(n)}\ [\rho_j\otimes \xi^{\otimes (n/2-j)}]\right)=\Tr\left(J_z^{(n)}\ \left[\frac{\beta_{j'|j}}{\Tr(\beta_{j'|j})} \otimes \xi^{\otimes (n/2-j')}\right]\right) \ .
\ee
Then, since the expectation value of angular momentum operators for $\xi$ is zero, then the two sides of the above equation simplify to
\be
\Tr(J^{(j)}_z \rho_j)= \Tr(J^{(j')}_z\frac{\beta_{j'|j}}{\Tr(\beta_{j'|j})})\ . 
\ee
Putting this into Eq.(\ref{cm}), we obtain the fidelity
\begin{align}
\text{Fid}(\mathcal{E})&=\frac{1}{2}+\sum_{j', j} \text{P}(j')\frac{\Tr(\beta_{j'|j} J_z)}{2j'}\\ &=\frac{1}{2}+\sum_{j'} \frac{1}{2j'}  \sum_j p_j\Tr(\beta_{j'|j})  \Tr(\rho_j J_z)\\ &= \frac{1}{2}+ \sum_{j',j} p_j\Tr(\beta_{j'|j}) \frac{j}{j'}  (f_j-1/2)\ ,
\end{align}
where to get the last line we have used Eq.(\ref{mb}). Finally, using
\be
\Tr(\beta_{j'|j})=\text{P}(j'|j,T)\ ,
\ee
we obtain Eq.(\ref{avgfid}).
\newpage

\newpage

\section{Coherent version of the protocol}\label{App:Coh}

Here, we describe a variant of the protocol that involves only unitary transformations and is therefore fully reversible. This makes it analogous to the full Schur transform and allows us to separate the information encoded in the irreps of $\mathrm{SU}(2)$ and $\mathbb{S}_n$, concentrating them in different systems.
 
First, consider the case of $n=2$ qubits.  Before the final measurement in the SWAP test circuit in Fig.\ref{fig:swap-test-box},  the output of the circuit  is
\be
\Big(H_{\text{anc}}\otimes\mathbb{I}\Big)\Big(|0\rangle\langle 0|_{\text{anc}}\otimes \mathbb{I}+|1\rangle\langle 1|_{\text{anc}}\otimes \text{SWAP}\Big) \big(|+\rangle_{\text{anc}}|\psi\rangle\big)=  |0\rangle_{\text{anc}} \otimes P^{(+)} |\psi\rangle +|1\rangle_{\text{anc}} \otimes P^{(-)} |\psi\rangle  \ ,
\ee
where $H_{\text{anc}}$ is the Hadamard gate on the ancilary qubit, and 
$P^{(+)}$ and $P^{(-)}$ are the projectors to the symmetric and the anti-symmetric subspaces, respectively.  This is indeed the Schur transform for $n=2$ qubits. 

For general $n$, in the original protocol discussed in the paper, at each round, a pair of qubits is randomly selected for a SWAP test. Moreover, these qubits are chosen based on the outcomes of the previous SWAP tests: if a pair of qubits is found in the singlet state, it is discarded and no longer used in subsequent SWAP tests. However, in the coherent version, since no measurements are performed, the outcomes of previous SWAP tests are not known. 

Therefore, the locations of the SWAP tests must also be chosen coherently. That is, in addition to the ancilla that records the outcome, we require an extra register to keep track of the locations of the qubits that are selected for each round of the SWAP test. These quantum registers allow us to track the positions of the detected singlets so that SWAP tests are not applied to them in subsequent rounds.

Therefore, the total Hilbert space is
\be
(\mathbb{C}^2)^{\otimes n} \otimes \mathcal{H}_{\text{ang-reg}}\otimes  \bigotimes_{t=1}^T(\mathcal{H}_{\text{loc},t}\otimes \mathcal{H}_{\text{out, t}}) \ , 
\ee
where $(\mathbb{C}^2)^{\otimes n}$ corresponds to the original $n$ qubits in the system, $\mathcal{H}_{\text{ang reg}}$ is the register that keeps track of the number of detected singlets (similar to the original algorithm) and has dimension $\lceil n/2 \rceil$. Furthermore, 
$T=2n\ln (n \epsilon^{-1})$ is the number of SWAP tests (a parameter of the algorithm), $\mathcal{H}_{\text{loc},t}$ stores the location of the $t$-th SWAP test, 
and $\mathcal{H}_{\text{out},t}$ is a single qubit that records the outcome of the SWAP test. The dimension of $\mathcal{H}_{\text{loc},t}$ is $\binom{n}{2}+1$.

As before, let $\{|j\rangle\}$ be an orthonormal basis for $\mathcal{H}_{\text{ang-reg}}$, where $n/2 - j$ is the number of detected singlets. This register is initialized in the state $|j = n/2\rangle$, as before.

At step $t = 1, \cdots, T$ of the algorithm, we apply a controlled unitary, controlled by $\mathcal{H}_{\text{ang-reg}}$, that prepares $\mathcal{H}_{\text{loc},t}$ in the uniform superposition
\be
|\Psi(j)\rangle = \frac{1}{\sqrt{\binom{2j}{2}}} \sum_{w=1}^{\binom{2j}{2}} |w\rangle\ , 
\ee
where $j$ is determined by the state of $\mathcal{H}_{\text{ang-reg}}$, and $\{|w\rangle : 0, \cdots, \binom{n}{2}\}$ is an orthonormal basis for $\mathcal{H}_{\text{loc},t}$. In particular, each value of $w$ in the above superposition uniquely labels a pair of qubits $(r, s): 1 \le r < s \le n$ from the $2j$ qubits in the $A$ part, which will be denoted as $w(r,s)$. The state $|\Psi(j)\rangle$ selects a pair of qubits, $r$ and $s$, uniformly at random, thereby determining the location of the next SWAP test.

To prepare the state $|\Psi\rangle_j$, we can apply  a controlled unitary  in the form
\be
\sum_{j} |j\rangle\langle j|_{\text{ang-reg}} \otimes \Big[ |\Psi(j)\rangle\langle 0|_{\text{loc},t} + |0\rangle\langle \Psi(j)|_{\text{loc},t} +\Pi_\perp\Big]\ ,
\ee
where $\Pi_\perp$ is the projector to the subspace of $\mathcal{H}_{\text{loc},t}$ that is orthogonal to $|0\rangle$ and $|\Psi(j)\rangle$. When this unitary is applied to the initial state $|j\rangle\otimes |0\rangle\in \mathcal{H}_{\text{ang-reg}}\otimes \mathcal{H}_{\text{loc},t}$, it produces  state $|j\rangle\otimes |\Psi(j)\rangle$.

In summary, at step $t$ of the algorithm, we implement a controlled unitary that depending on the state of $\mathcal{H}_{\text{ang reg}}$  prepares $\mathcal{H}_{\text{loc},t}$ in state $|\Psi(j)\rangle $, and then apply  the SWAP test on pair of qubits $(r,s)$, which can be realized by the following unitary on the original $n$ qubits in the system, and  $\mathcal{H}_{\text{loc}, t}$ and $\mathcal{H}_{\text{out}, t}$:  
\be
V_t=|0\rangle\langle 0|_{\text{out}, t}\otimes \mathbb{I}\ +\   |1\rangle\langle 1|_{\text{out}, t}\otimes  \left(\sum_{r,s: 1\le r<s\le n}|w(r,s)\rangle\langle w(r,s)|_{\text{loc}, t} \otimes \text{SWAP}_{r,s}\right) \ ,
\ee
where $\text{SWAP}_{r,s}$ denotes the SWAP gate performed on the pair $(r,s)$ of qubits in the system and an ancilla qubit. 

Now applying this unitary on the system in arbitrary state $|\psi\rangle\in (\mathbb{C}^2)^{\otimes n}$, and the ancillary qubit with the Hilbert space $\mathcal{H}_{\text{out}, t}$ in state  $|+\rangle=(|0\rangle+|1\rangle)/\sqrt{2}$, and then applying Hadamard on  the ancilla qubit $\mathcal{H}_{\text{out}, t}$, we obtain 
\be
(H_{\text{out}, t}\otimes \mathbb{I})V_t\Big( |\Psi(j)\rangle \otimes   |+\rangle_{\text{out}, t} \otimes |\psi\rangle\Big)=\frac{1}{\sqrt{{2j}\choose{2}}} \sum_{r<s}\   |w(r,s)\rangle_{\text{loc}, t}\otimes \left[ |0\rangle_{\text{out}, t}\otimes P_{r,s}^{(+)}|\psi\rangle+ |1\rangle_{\text{out}, t}\otimes P_{r,s}^{(-)}|\psi\rangle  \right]\ .
\ee
Now, in the original version of the protocol, we measure the ancilla qubit, and if it is found in state $1$, we decrease the value of the angular momentum register as $|j\rangle\rightarrow |j-1\rangle$, and then move the qubits $r$ and $s$ to batch $B$. Here, we can do this in a controlled fashion. I.e, we apply a controlled subtraction that decreases the value of the 
register as $|j\rangle \rightarrow |j-1\rangle$, which can be realized using a cyclic shift controlled by the ancilla. Furthermore,  if the ancilla qubit is in state $|1\rangle_{\text{out}, t}$,    we SWAP qubits $r,s$ with qubits $2j$ and $2j-1$. This can again be done with a controlled unitary, controlled by the ancilla qubit $\text{out}, t$. Then, we go to the next round of the algorithm, i.e., apply the SWAP test $t+1$ with a similar protocol.  

Then, while the entire process is unitary, it can be easily seen that after
$T$ SWAP tests the reduced state of
$(\mathbb{C}^2)^{\otimes n} \otimes \mathcal{H}_{\text{ang-reg}}$, is identical with the reduced state obtained from the original version of the algorithm, i.e., state
\be
\mathcal{E}^T(|\psi\rangle\langle\psi|\otimes |n/2\rangle\langle n/2|)\ .
\ee
This can be seen by noting that if 
the discarded systems
$\mathcal{H}_{\text{loc},t} : t=1,\cdots, T$ are all measured in the  $\{|w\rangle\}$ basis, and  ancillary qubits 
$\mathcal{H}_{\text{out, t}}: t=1,\cdots, T$ are all measured in $\{|0\rangle,|1\rangle\}$ basis, then the realized protocol is identical with the original protocol. It follows that the reduced state of $(\mathbb{C}^2)^{\otimes n} \otimes \mathcal{H}_{\text{ang-reg}}$ is fully independent of the original reduced state of irreps of $\mathbb{S}_n$, namely subsystems $\mathbb{C}^{m(n,j)}$ in Eq.(\ref{dec}). Given that the entire process is unitary and the reduced state of $(\mathbb{C}^2)^{\otimes n} \otimes \mathcal{H}_{\text{ang-reg}}$ is independent of these subsystems, this means that the information originally encoded in these subsystems should be recoverable from the complimentary subsystems, namely $\bigotimes_{t=1}^T(\mathcal{H}_{\text{loc},t}\otimes \mathcal{H}_{\text{out, t}})$. This can be formulated, e.g., in terms of Uhlmann's theorem, or Information-Disturbance Tradeoff (See, e.g., \cite{kretschmann2008information}).

\end{document}